\documentclass[a4paper]{aa}  
\usepackage{graphicx}
\usepackage{longtable}
\usepackage{txfonts}

\usepackage{natbib}

\begin{document}
\setcounter{table}{0}
\setcounter{figure}{0}

\title{Hot Jupiters and the evolution of stellar angular momentum}

    \titlerunning{Exoplanets and stellar angular momentum evolution}
    \authorrunning{A. F. Lanza}


   \author{A.~F.~Lanza}

   \offprints{A.~F.~Lanza}

   \institute{INAF-Osservatorio Astrofisico di Catania, Via S. Sofia, 78 
               -- 95123 Catania, Italy \\ 
              \email{nuccio.lanza@oact.inaf.it}    
             }

   \date{Received ... ; accepted ... }

    \abstract{Giant planets orbiting  main-sequence stars closer than 0.1 AU are called hot Jupiters. They interact with their stars affecting {{ their angular momentum}}.}{Recent observations provide suggestive evidence of excess angular momentum in stars with hot Jupiters in comparison to stars with distant and less massive planets. This has been  attributed to tidal interaction, but needs to be investigated in more detail considering also other possible explanations {{ because in several cases the tidal synchronization time scales are much longer than the ages of the stars.}}}{{{ We select  stars harbouring transiting hot Jupiters to study their rotation and find that those with an effective temperature $T_{\rm eff} \ga 6000$~K and a rotation period $P_{\rm rot} \la 10$ days are  synchronized with the orbital motion of their planets or have a rotation period approximately twice that of the planetary orbital period. Stars with $T_{\rm eff} \la 6000$~K or $P_{\rm rot} \ga 10$ days show a general trend toward synchronization with increasing effective temperature or decreasing orbital period. We propose  a model for the angular momentum evolution of stars with hot Jupiters to interpret these observations.}} It is based on the hypothesis that a close-in giant planet affects the  coronal field of its host star leading to a topology with predominantly closed field lines. {{ An analytic linear force-free model is adopted to compute the radial extension of the corona and its angular momentum loss rate. }}The corona is more tightly confined in F-type stars, and in G- and K-type stars with a rotation period shorter than $\sim 10$ days. The  angular momentum loss is produced by coronal eruptions similar to solar coronal mass ejections.}{{{ The model predicts that F-type stars with hot Jupiters, $ T_{\rm eff} \ga 6000$~K and an initial rotation period $ \la 10$ days suffer no or very little angular momentum loss during their main-sequence lifetime. This can explain their rotation as a remnant of their pre-main-sequence evolution. On the other hand, F-type stars with $P_{\rm rot} > 10$ days, and G- and K-type stars experience a significant angular momentum loss during their main-sequence lifetime, but at a generally slower pace than similar stars without close-in massive planets. Considering a spread in their ages, this can explain the observed rotation period distribution of  planet-harbouring stars.}}}{Our model can be tested observationally and has relevant consequences for the relationship between stellar  rotation and close-in giant planets as well as for the application of  gyrochronology to estimate the age of planet-hosting stars.}

   \maketitle


\section{Introduction}
\label{intro}

The search for planetary systems has revealed a population of planets having a mass comparable to that of Jupiter orbiting main-sequence late-type stars closer than $\sim 0.1$ AU\footnote{http://exoplanet.eu/}. They include about 25 percent of all known planets and have been called hot Jupiters because of their close proximity to their host stars. Remarkable  interactions between those planets and their host stars are expected,  both as a consequence of tides or  reconnection between planetary and stellar magnetic fields \citep[cf., e.g., ][]{Cuntzetal00,GuSuzuki09}. Evidence of a magnetic interaction { was first shown by \citet{Shkolniketal03} } and  has been recently reviewed by, e.g., \citet{Lanza08}, \citet{Lanza09}, and \citet{Shkolniketal09}.

In the present study, we focus on the modifications induced in a stellar corona by a close-in giant planet and their consequences for the evolution of the stellar angular momentum.  
{ The presence of a close-in giant planet may significantly affect the structure and the energy balance of the coronal field. \citet{Kashyapetal08} show that stars with hot Jupiters have X-ray luminosities up to $3-4$ times greater than  stars with distant planets. This suggests that a close-in planet may  enhance magnetic energy dissipation or lead to a predominance of closed and brighter magnetic structures in a stellar corona.
}

\citet{Pont09} has found that the sample of stars with transiting hot Jupiters show a statistic excess of rapidly rotating objects in comparison to stars without close-in planets. Plotting stellar rotation rate vs. the orbital semimajor axis, normalized to the average of the stellar and planetary radii, and the planet-to-star mass ratio, he finds some empirical evidence of faster rotation in stars with closer and more massive planets. This may be interpreted as an indication of tidal interaction that drives stellar rotation toward synchronization with the planetary orbital period because the tidal torque  is expected to increase for closer and more massive planets 
{ \citep[e.g., ][]{MardlingLin02}}.

There are a few examples of stars whose rotation appears to be synchronized with the orbit of their close-in planets, notably \object{$\tau$ Bootis} \citep[][]{Donatietal08}, and the transiting system \object{CoRoT-4} \citep{Aigrainetal08,Lanzaetal09b}.
They are F-type stars with a shallow outer convective envelope. 
{  In particular, $\tau$~Boo has an estimated age of $\approx 2$~Gyr and is orbited by a planet 
with $ M \sin i = 4.38$ M$_{J}$ corresponding to 
$\approx 6.8$ Jupiter masses at a distance of $\sim 0.049$~AU, i.e., $\sim 7.2$ stellar radii, if an inclination $ i \simeq 40^{\circ}$ is adopted, as suggested by stellar Doppler Imaging models  \citep{Leighetal03,Catalaetal07}. 
} 
Assuming that only the envelope of $\tau$ Boo is in a synchronous rotation state, \citet{Donatietal08} found a synchronization timescale compatible with the main-sequence lifetime of the star. However, in the case of CoRoT-4a that timescale is of the order of 350 Gyr, because the semimajor axis of the planetary orbit is 17.4  stellar radii and the mass of the planet is only 0.72 Jupiter masses making any tidal interaction extremely small \citep{Lanzaetal09b}. Therefore, a different process is required to account for the synchronization of CoRoT-4a. Two other intriguing cases are those of the host stars of \object{XO-4} and \object{HAT-P-6} whose rotation periods are close to twice the orbital periods of their transiting planets, respectively \citep{McCulloughetal08}. 

Tidal interactions are not the only processes affecting spin and orbital angular momenta.
Since late-type stars have magnetized stellar winds that produce a remarkable braking of their  rotation during their main-sequence lifetime, a continuous loss of angular momentum  must be taken into account to model the evolution of the stellar spin  in a proper way. As shown by, e.g.,  \citet{DobbsDixonetal04}, this may have a significant impact not only on  the evolution of stellar rotation but also on that of the orbital parameters,
{ notably the eccentricity and the semimajor axis,} especially during the initial stages of the main-sequence evolution when the star is a fast rotator and stores most of the angular momentum of the system { that can be transferred to the planetary orbit to excite its eccentricity and/or increase its semimajor axis}. 

\citet{Lanza08}  proposed a model for the interaction between the coronal and the planetary magnetic fields considering a linear force-free equilibrium for the coronal  field. Here we shall  apply that model to study the angular momentum loss from the coronae of stars hosting hot Jupiters assuming that a close-in planet leads to a corona with predominantly closed magnetic field lines 
{ because it tends to reduce the magnetic helicity of the field via a steady dissipation of magnetic energy associated with its motion through the corona \citep[cf. ][]{Lanza09}. Fields with a lower helicity are characterized by a topology with a greater fraction of closed field lines, while an increase of the helicity beyond a certain threshold may lead to an opening up of the field lines \citep{Zhangetal06,ZhangFlyer08}. The recent MHD simulations of the effects of a close-in massive planet on the coronal field of a star by \citet{Cohenetal09} confirm that the planet inhibits the expansion of the coronal field and the acceleration of the stellar wind.  

Under the hypothesis of a coronal field with predominantly closed field lines}, we find that a close-in planet may significantly reduce magnetic braking in rapidly rotating stars. This, in addition to tidal interaction, may explain the tendency toward synchronization found by \citet{Pont09} if the initial rotation periods of the stars are close to the orbital periods of their planets, in agreement with models of rotational evolution  which assume that both a star and its planet are dynamically coupled to a circumstellar disc during the first few million years of their evolution (cf. Sect.~\ref{pms_evolution}). 

\section{Properties of transiting planetary systems}
\label{observations}

\begin{figure*}[t]
\centerline{\includegraphics[width=16cm,height=18cm,angle=90]{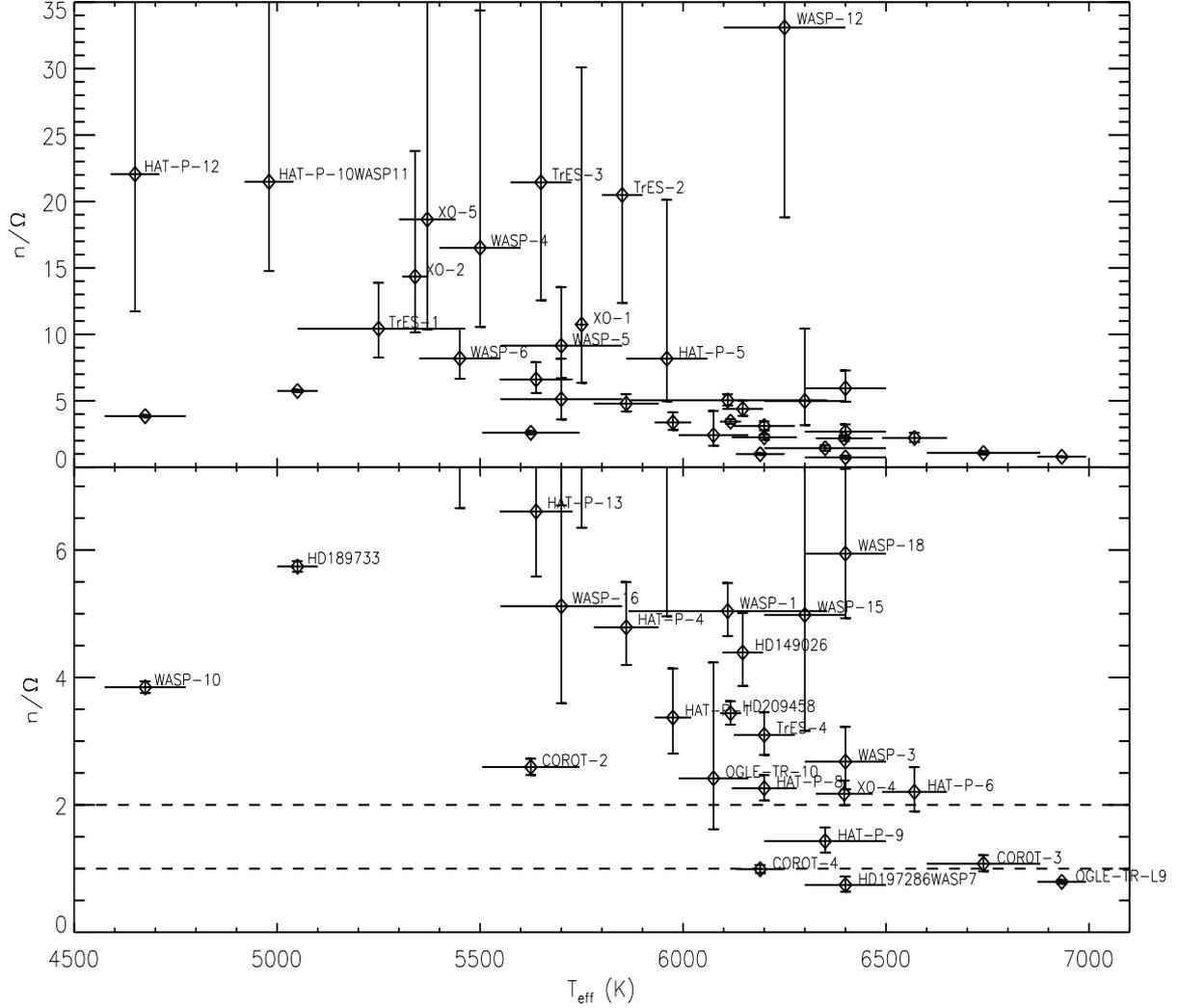}}
\caption{Upper panel: The synchronization parameter $n/\Omega$ vs. the effective temperature of the star in transiting  planetary systems; lower panel: an enlargement of the lower portion of the upper panel, to better show the correlation close to  $n/\Omega=1$ and  $n/\Omega = 2$, which are marked by horizontally dashed lines. The names of the systems are reported in both panels, although they are omitted for $n / \Omega \leq 7.5$ in the upper panel to avoid confusion.  }
\label{teff_corr}
\end{figure*}
\begin{figure*}[t]
\centerline{\includegraphics[width=10cm,height=18cm,angle=90]{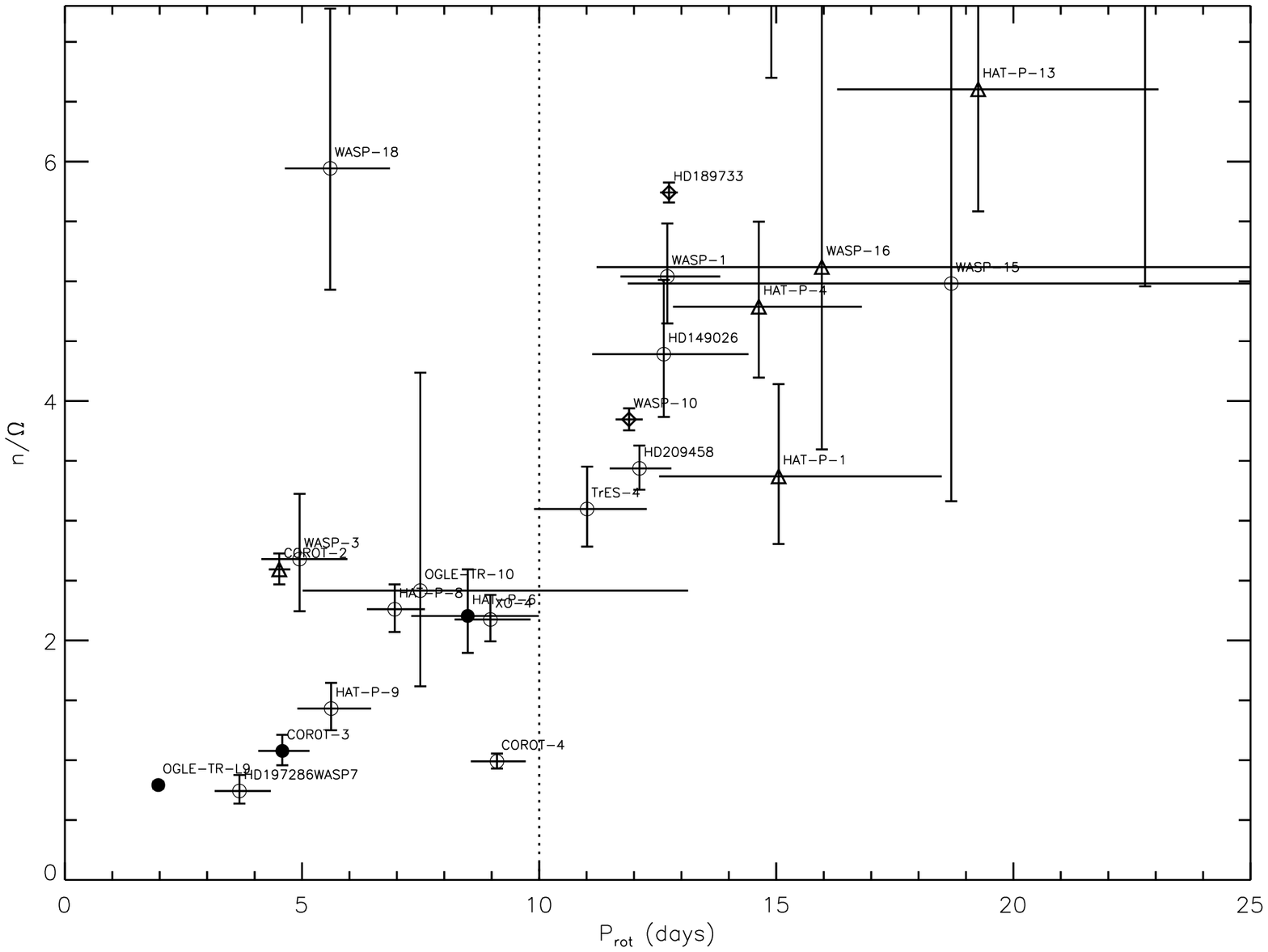}}
\caption{The synchronization parameter $n/\Omega$ vs. the stellar rotation period $P_{\rm rot}$ for our sample of transiting extrasolar planetary systems. Different symbols indicate different ranges of stellar effective temperatures: filled circles: $T_{\rm eff} \geq 6500$~K; open circles: $6000 \leq T_{\rm eff} < 6500$~K; triangles: $5500 \leq T_{\rm eff} < 6000$~K. 
The vertically dotted line marks the rotation period of 10 days below which systems show a tendency to cluster around the values $n/\Omega =1$ (synchronization) and $n/\Omega =2$.
   }
\label{synch_vs_prot}
\end{figure*}

The rotation periods of planet-hosting stars are difficult to measure from the rotational modulation of their flux because planets can be detected by radial velocity monitoring only around relatively inactive stars, i.e., with a photometric modulation below $\approx 0.01$~mag. The situation is going to change with space-borne photometry made possible by CoRoT and Kepler \citep[cf., e.g., ][]{Alonsoetal08,Aigrainetal08}, but at the moment the most reliable measurements of the rotation rates of planet-harbouring stars come from spectroscopy, viz., from rotational line broadening $v \sin i$. To find the rotation velocity $v$, we need to know the inclination of the rotation axis $i$, which can be derived in the case of transiting systems by assuming that the stellar and orbital angular momenta are aligned. This alignment results from the formation of planets in a circumstellar disc, if the gravitational interaction among planets is not too strong, otherwise one expects a significant misalignment of the spin and orbital angular momenta as well as eccentric planetary orbits \citep[e.g., ][]{Nagasawaetal08}. It is interesting to note that the angle $\lambda$ between the projections of the spin and orbital angular momenta on the plane of the sky can be measured through the Rossiter-McLaughlin effect that has already been detected in several systems \citep[][]{Ohtaetal05,FabryckyWinn09}.  

In view of the possibility of guessing the value of the inclination, we focus on transiting systems. As to the end of September 2009, 62 transiting systems are known\footnote{See~http://exoplanet.eu/}. We esclude those with an eccentricity  $e \geq 0.06$ and/or with a significant misalignment as indicated by the Rossiter-McLaughlin effect. { They are listed in Table~\ref{table_excluded}  where we report, from the first to the fourth column, the name of the system, its orbital eccentricity, the misalignment angle $\lambda$ as derived from the Rossiter-McLaughlin effect, and the references, respectively. }
                           \begin{table*}
               \begin{tabular}{lccl}
               Name  & $e$ & $\lambda$ & References \\
                  & & & \\
               HD~17156 & $0.6819 \pm 0.0044$ & & \citet{Barbierietal09} \\
               HD~80606 & $0.9332 \pm 0.0008$ & & \citet{FisherValenti05,Pontetal09} \\
               HD~147506/HAT-P-2 & $0.5163 \pm 0.025$ & & \citet{Bakosetal07,Loeilletetal08} \\
               GJ~436 & $0.160 \pm 0.019$ & & \citet{Beanetal08,Demoryetal07,Manessetal07}\\
               XO-3 & $0.2884 \pm 0.0035$ & $37^{\circ} \pm 4^{\circ}$ & 
                               \citet{Johns-Krulletal08,Winnetal09} \\
               HAT-P-7 & & $182^{\circ} \pm 10^{\circ}$ & \citet{Naritaetal09,Winnetal09a} \\
               HAT-P-11 & $0.198 \pm 0.046$ & & \citet{Dittmannetal09,Bakosetal09b} \\
               WASP-14 & $0.091 \pm 0.003$ & $-33^{\circ} \pm 8^{\circ}$ & \citet{Joshietal09,Johnsonetal09a} \\
               WASP-17 & $0.129 \pm 0.1$ & $-147^{\circ} \pm 50^{\circ}$ & \citet{Andersonetal09} \\
               CoRoT-1 & & $77^{\circ} \pm 11^{\circ}$ & \citet{Pontetal09} \\
                   & & &  \\ 
                           \end{tabular}
\caption{Transiting planetary systems excluded from the present analysis.}
\label{table_excluded}
                             \end{table*}
 Moreover, we exclude CoRoT-7 because its planets are not hot Jupiters, but terrestrial-sized objects \citep[][]{Legeretal09,Quelozetal09}; CoRoT-6 because it was announced, but its parameters have not been published yet; \mbox{WASP-2}, \mbox{WASP-8}, \mbox{OGLE-TR-111}, \mbox{OGLE-TR-182}, and \mbox{OGLE-TR-211}  because no data on their stellar rotation were found in the literature. 

The relevant parameters of the considered systems are listed in Tables~\ref{table_p1} and \ref{table_p2}. In Table~\ref{table_p1},  from the first to the eighth column, we list the system name, the orbital period $P_{\rm orb}$, the semimajor axis of the planetary orbit $a$, the radius of the planet $R_{\rm p}$, the mass of the planet $M_{\rm p}$, the radius of the star $R$, the mass of the star $M$, and its effective temperature $T_{\rm eff}$, respectively; in Table~\ref{table_p2},  from the first to the seventh column, we list the system name, the $v \sin i$ of the star, the rotation period of the star as derived from the rotational modulation of its flux $P_{\rm rot}$, the minimum and maximum estimated age for the star, the synchronization time for stellar rotation $\tau_{\rm sync}$, the  eccentricity of the planetary orbit $e$, and the references for both Tables~\ref{table_p1} and \ref{table_p2}, respectively. When it is not determined from the rotational modulation of the flux, we estimate the 
rotation period of a star as: $P_{\rm rot} = 2\pi R/(v \sin i)$, where the radius of the star 
is taken from Table~\ref{table_p1} and an inclination $i =90^{\circ}$ is adopted, which is perfectly adequate for transiting systems in view of the average precision of our $v \sin i$ and stellar radii. In this case, we do not list $P_{\rm rot}$ in Table~\ref{table_p2}.
We make use of the formulae of \citet{MardlingLin02} to compute $\tau_{\rm sync}$ as:
\begin{equation}
\tau_{\rm sync}^{-1} \equiv \frac{1}{\Omega} \left| \frac{d\Omega}{dt} \right| = \frac{9}{2} \frac{1}{\gamma^{2} Q^{\prime}_{*}} \left( \frac{M_{\rm p}}{M} \right)^{2} \left( \frac{R}{a} \right)^{9/2} \left| 1 - \left( \frac{n}{\Omega} \right) \right| \sqrt{\frac{G M}{R^{3}}},
\label{tidal_time}
\end{equation}
where $\gamma R \simeq 0.35\, R$ is the gyration radius of the star, $Q_{*}^{\prime}$ is the modified tidal quality factor of the star defined as 
$Q_{*}^{\prime} \equiv 3 Q_{*}/2k_{*}$, where $Q_{*}$ is the specific dissipation function and $k_{*}$ is the Love number \citep[cf., e.g., ][]{MurrayDermott99};  $n\equiv 2\pi / P_{\rm orb}$ is the mean orbital motion,  $\Omega \equiv 2\pi / P_{\rm rot}$ the angular velocity of rotation of the star, and $G$ the gravitation constant. Eq.~(\ref{tidal_time}) is valid for circular orbits and when the spin axis is aligned with the orbital angular momentum. To compute the synchronization times listed in Table~\ref{table_p2}, we adopt $Q^{\prime}_{*} = 10^{6}$ and assume that the entire star is synchronized as it is customary in tidal theory and is suggested by the  tidal evolution of close binaries  observed in stellar clusters of different ages. 
If $n=\Omega$, i.e., when the stellar rotation is synchronized with the orbital motion, there is no tidal contribution to the variation of the stellar angular momentum.
If we consider the case of F-type stars discussed in the Introduction, i.e., $\tau$ Boo and CoRoT-4a,  since their initial rotation is synchronized with the orbital motion, tidal effects can be neglected. Nevertheless, we list in Table~\ref{table_p2} a lower limit for the synchronization timescale for CoRoT-4a  computed by assuming $\Omega = 2 n$. 

We note that $\tau_{\rm sync}$ regards the angular momentum exchange between the 
orbital motion and the stellar spin. On the other hand, orbital circularization proceeds much faster owing to the dissipation of energy inside the planet at almost constant orbital angular momentum. The dissipation of energy inside the star is smaller by a factor of $10^{-2}-10^{-3}$  when we adopt $Q^{\prime}_{\rm p} = 10^{5}$ for the planet and $Q_{*}^{\prime}=10^{6}$  \citep[cf., e.g., ][]{Matsumuraetal08}. 

As a measure of  synchronization of a planet-harbouring star, we adopt the ratio  
 $ n/\Omega \equiv P_{\rm rot}/P_{\rm obs} $. 
Its dependence on the orbital parameters has already been investigated by, e.g., \citet{Pont09} and \citet{Levrardetal09}. Here, we focus on the correlation between $n/\Omega$ and the effective temperature of the star that is plotted in Fig.~\ref{teff_corr}.  Although there is a large scatter for $T_{\rm eff} \la 6000$~K, there is a  general trend toward synchronization, i.e., a decrease of $n/\Omega$ toward unity, with increasing effective temperature and the synchronized systems have $T_{\rm eff} \geq 6200$ K. We note that for $ 6000 \leq T_{\rm eff} < 6500$ K there is still a significant scatter in $n / \Omega$. Nevertheless, two subgroups of systems can be identified:  one consists of those being close to $n / \Omega  =1 $ or to $n/\Omega = 2$, while the other consists of those showing $n / \Omega $ remarkably greater than 2. These two groups appear to be separated if we plot $ n/\Omega$ vs. $P_{\rm rot}$, as it is shown in Fig.~\ref{synch_vs_prot}, with all systems with $P_{\rm rot} \leq 10$ days having $n /\Omega \la 2$, with the exception of WASP-18.

\begin{figure}[t]
\centerline{\includegraphics[width=8cm,height=12cm,angle=0]{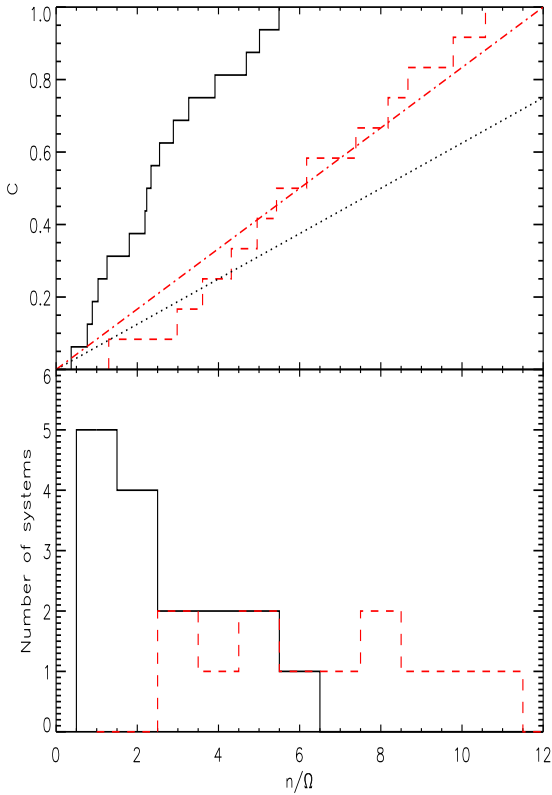}}
\caption{{ Upper panel:} Cumulative distribution functions $C$ of $n/\Omega$ for the systems with a stellar effective temperature $T_{\rm eff} \geq 6000 $~K (black solid line) and $T_{\rm eff} < 6000$~K (red dashed line) and $n/\Omega \leq 12$; the expected cumulative distributions for a uniform distribution of $n/\Omega$ are also plotted for $T_{\rm eff} \geq 6000$~K (black dotted line) and $T_{\rm eff} < 6000$~K (red dot-dashed line), respectively. { Lower panel:} Histogram of the distribution of $n/\Omega$ for systems with $T_{\rm eff} \geq 6000$~K (black solid line) and $T_{\rm eff} < 6000$~K (red dashed line), respectively.    }
\label{KS_distr}
\end{figure}

\begin{figure}[t]
\centerline{\includegraphics[width=9cm,height=7cm,angle=0]{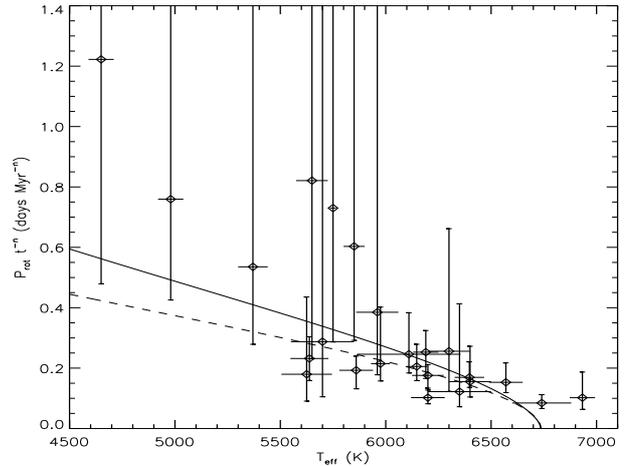}}
\caption{Age-normalized rotation period, $P_{\rm rot} t^{-n}$ with $n=0.5189$, vs. the stellar effective temperature $T_{\rm eff}$ for our sample of stars with transiting hot Jupiters. The relationship found by \citet{Barnes07} for  stars without hot Jupiters is plotted with a solid line. The dashed line plots the same relationship proposed by Barnes but for a lower angular momentum loss rate parameter  $a=0.56$.}
\label{prot_teff}
\end{figure}

{ The significance of the clustering of the systems around $n/\Omega = 1, \, 2$ can be assessed by means of the Kolmogorov-Smirnov test (hereinafter KS)  based on the cumulative distribution function of the observed values \citep[see, e.g., ][]{Pressetal92}. In the upper panel of Fig.~\ref{KS_distr}, we compare the cumulative distribution functions for the systems with $T_{\rm eff} \geq 6000$~K and $T_{\rm eff } < 6000$~K with  the corresponding uniform distributions for $n/\Omega \leq 12$, respectively. The probability that the observed distribution of $n /\Omega$ for $T_{\rm eff} \geq 6000$~K is drawn from a uniform distribution is only $2.50 \times 10^{-6}$, according to the KS statistics. On the other hand, the probability that the distribution of $n/\Omega$  for $T_{\rm eff} < 6000$~K is compatible with a uniform distribution is $0.575$. In the lower panel of Fig.~\ref{KS_distr}, we plot an histogram with bin sizes of unity  to better show the clustering of $n/\Omega$ around the values $\leq 2$ for $T_{\rm eff} \geq 6000$~K and the almost uniform distribution  for $T_{\rm eff } < 6000$~K. In Table~\ref{KS_test}, we list the results of the KS test for different maximum values of $n/\Omega$ to show that the clustering around the values $1$ and $2$ for $T_{\rm eff} \geq 6000$~K is significant and does not depend on the choice of the upper limit of $n/\Omega$ used to define the data sample. Specifically, in the columns from the first to the fifth, we list the maximum value of $n/\Omega$ defining the  sample, the probability $P_{1}$ that the subset with $T_{\rm eff} \geq 6000$~K is drawn from a uniform distribution, the number $N_{1}$ of systems in that subset, the probability $P_{2}$ that the subset with $T_{\rm eff} < 6000$~K is drawn from a uniform distribution, and the number $N_{2}$ of systems in that subset, respectively. For subsets  containing less than $6$ systems, we do not give the corresponding probability because the KS statistics is computed by an asymptotic formula which is not accurate in these cases \citep[cf. ][]{Pressetal92}. 

The trend toward synchronization seen in Fig.~\ref{teff_corr} could be due to the general decrease of the rotation periods of main-sequence stars with increasing effective temperature and the clustering of the orbital periods of the hot Jupiters around $\sim 3-4$ days, irrespective of any kind of star-planet interaction. To test this explanation, we compare the evolution of the rotation periods of planet-hosting stars with those of the stars without planets as parameterized by \citet{Barnes07} from the study of stellar rotation in open clusters and in the field. It is important to consider the stellar age in addition to the effective temperature because stellar rotation periods on the main sequence show a remarkable dependence on both those parameters.
According to \citet{Barnes07}, the rotation period $P_{\rm rot}$ in days and the age $t$ in Myr are related to the $B-V$ colour index according to the formula:
\begin{equation}
P_{\rm rot} t^{-n} = a [(B-V) - 0.4]^{b},
\label{braking_law}
\end{equation}
where $n = 0.5189 \pm 0.007$ is almost identical to the value $1/2$ of the well-known Skumanich braking law, $a=0.7725 \pm 0.011$ and $b=0.601 \pm 0.024$ \citep[cf. Eq.~(3) in ][]{Barnes07}. It is interesting to note that Eq.~(\ref{braking_law}) with $n =1/2$, i.e., the Skumanich's law, can be obtained by assuming an  angular momentum loss rate: $d L / dt = K(B-V) \, \Omega^{3}$, where  $K \propto a  [(B-V) - 0.4]^{b}$ depends on the parameters $a$ and $b$ of the Barnes' relationship. 

In view of the dependence on the stellar age, we restrict our comparison to the 24 systems with an age estimate in the literature and plot in Fig.~\ref{prot_teff} their $P_{\rm rot} t^{-n}$ vs. $T_{\rm eff}$, where $n=0.5189$ and the effective temperature has been converted into the $B-V$ colour index thanks to the calibration by 
\citet{Bessell79}. The solid line is the correlation found by \citet{Barnes07} for stars without hot Jupiters. The two systems with $T_{\rm eff} > 6700$~K are CoRoT-3 and OGLE-TR-L9 whose rotation periods are significantly affected by tidal interaction (see below), so they are excluded from our analysis.

A $\chi^{2}$ test of the goodness of fit of the Barnes' relationship for the remaining 22 systems has been performed and gives a probability of 0.14 that the obtained $\chi^{2}=28.96$ is compatible with  Eq.~(\ref{braking_law}), suggesting that stars with transiting hot Jupiters are, on the average, faster rotators than similar stars without planets of the same age. The goodness-of-fit probability decreases to $0.018$ if we restrict the comparison to the 15 stars with $5500 \leq T_{\rm eff} < 6700$~K ($\chi^{2} = 28.61$), reinforcing the  conclusion for this subsample of stars.

On the other hand,  assuming $a=0.56$, we obtain $\chi^{2}=12.66$ for the 22 systems with $T_{\rm eff} < 6700$~K, which has a goodness-of-fit probability of 0.94, indicating that a Skumanich-type law with a reduced angular momentum loss rate  is an adequate description of the overall evolution of the rotation of stars with hot Jupiters. This implies that the rotation periods of the planet-hosting stars are, on the average, a factor of 0.7 shorter than those of the stars without planets of the same age. This is particularly significant  because the detection of transits and the radial velocity confirmation of planets introduce a bias toward stars with a lower level of activity and therefore longer rotation periods. An important consequence of the lower angular momentum loss rate found in stars with hot Jupiters is  that the tendency toward synchronization shown in Fig.~\ref{teff_corr} cannot be explained on the basis of the general  decrease of the rotation period with effective temperature observed in main-sequence stars without hot Jupiters. 
}

\begin{figure*}[t]
\centerline{\includegraphics[width=10cm,height=20cm,angle=90]{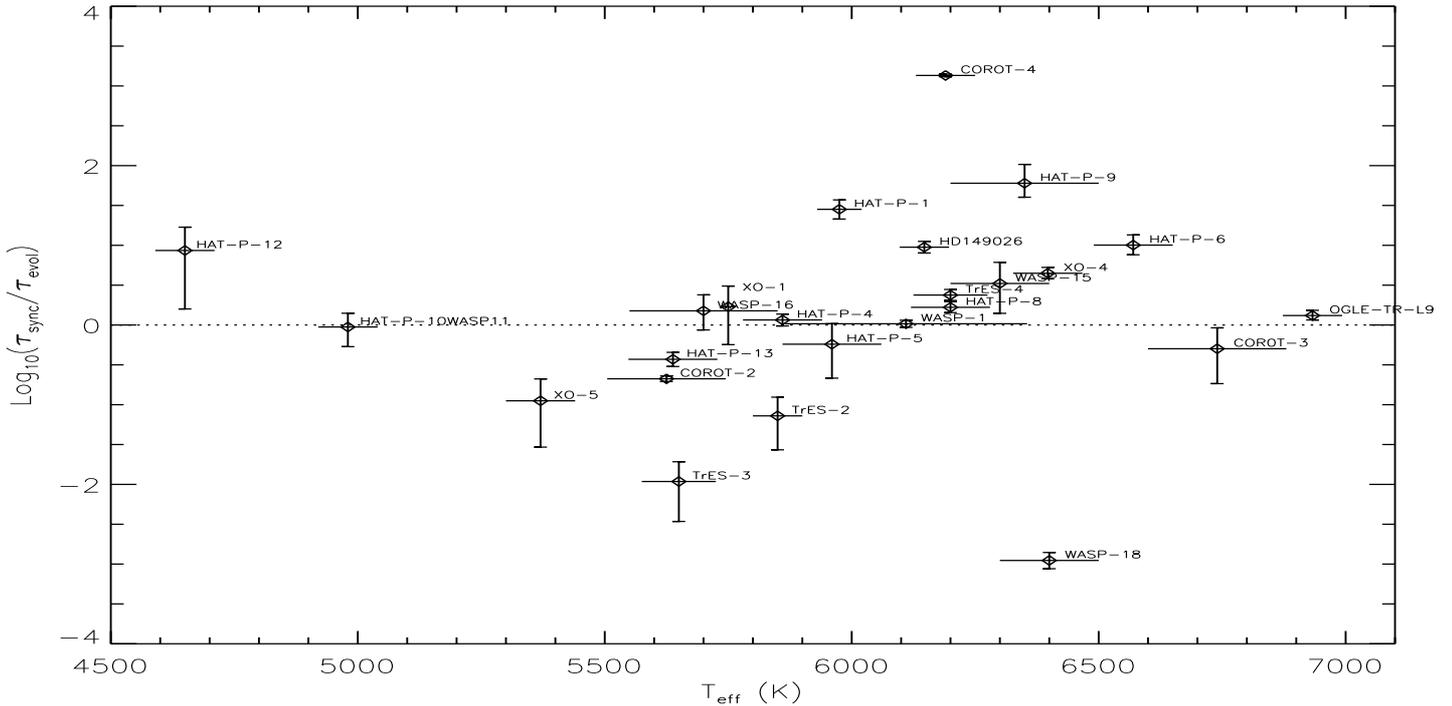}}
\caption{The ratio between the tidal synchronization timescale $\tau_{\rm sync}$ and an estimated upper limit for the age of the star $\tau_{\rm evol}$ vs. the effective temperature for the systems  for which an estimate of the stellar age has been found in the literature (cf. Table~\ref{table_p2}).   }
\label{teff_tau_corr}
\end{figure*}

The general trend shown in Fig.~\ref{teff_corr} as well as the correlation in Fig.~\ref{synch_vs_prot}, may result from tidal effects, but a plot of the ratio $\tau_{\rm sync}/\tau_{\rm evol}$, where $\tau_{\rm evol}$ is an estimated upper bound for the stellar age, vs. the effective temperature, shows a general increase of that ratio with the effective temperature (cf. Fig.~\ref{teff_tau_corr}). Morover, $\sim 70$ percent of the systems for which we have an age estimate have not had enough time to synchronize the rotation of their stars. The value of $\tau_{\rm sync}$ is  one or two orders of magnitude greater than the maximum estimated stellar age in the case of CoRoT-4, HAT-P-9 ($n/ \Omega \simeq 1$),  HAT-P-6, or XO-4 ($n/ \Omega \simeq 2$). Even by decreasing the value of $Q_{*}^{\prime}$ by one order of magnitude, we cannot eliminate the discrepancy. Therefore, tidal effects alone do not appear to be a viable explanation for the synchronization observed in several systems. 
A similar conclusion is reached for the correlation between $n /\Omega$ and $P_{\rm rot}$ seen in Fig.~\ref{synch_vs_prot} because systems such as CoRoT-4, HAT-P-6, HAT-P-9, or XO-4 have too long synchronization time scales  to explain their low values of $n /\Omega$. 
This implies that another mechanism must be at work, in addition to tidal effects, to produce the observed trend toward synchronization with increasing effective temperature and the remarkable concentration of systems around the values $n/\Omega = 1$ and $n/\Omega =2$ observed for $P_{\rm rot} \leq 10$ days and $T_{\rm eff} \geq 6000$~K. We conjecture that such a mechanism is related to the modification of the stellar coronal field and to the different magnetically-controlled angular momentum loss induced by a close-in massive planet. In the framework of such a hypothesis, we introduce in the next Sections a model for the angular momentum content of the corona of a star harbouring a hot Jupiter and discuss how the evolution of its angular momentum is affected.


                           \begin{table*}
               \begin{tabular}{lccccccc}
                       & & & & & & & \\ 
       Name & $P_{\rm orb}$ & $a$ & $R_{\rm p}$ & $M_{\rm p}$ & $R$  & $M$ & $T_{\rm eff}$ \\ 
                    & (days) & (AU) & ($R_{\rm J}$) & ($M_{\rm J}$)  & ($R_{\odot}$) & ($M_{\odot}$) & (K) \\ 
                       & & & & & & & \\ 
        \object{HD149026}  &  2.87589  &  0.04320  &  0.728 $\pm $  0.012  &  0.360 $\pm $  0.030  &  1.497 $\pm $  0.069  &  1.300 $\pm $    0.060  &    6147 $\pm $   50   \\ 
        \object{HD189733}  &  2.21858  &  0.03100  &  1.154 $\pm $  0.017  &  1.150 $\pm $  0.040  &  0.755 $\pm $  0.011  &  0.820 $\pm $    0.030  &    5050 $\pm $   50   \\ 
   \object{HD197286/WASP7}  &  4.95466  &  0.06180  &  0.925 $\pm $  0.043  &  0.960 $\pm $  0.200  &  1.236 $\pm $  0.050  &  1.280 $\pm $    0.160  &    6400 $\pm $  100   \\ 
        \object{HD209458}  &  3.52475  &  0.04700  &  1.320 $\pm $  0.025  &  0.657 $\pm $  0.006  &  1.125 $\pm $  0.022  &  1.101 $\pm $    0.064  &    6117 $\pm $   26   \\ 
          \object{TrES-1}  &  3.03007  &  0.03930  &  1.081 $\pm $  0.029  &  0.760 $\pm $  0.050  &  0.811 $\pm $  0.020  &  0.890 $\pm $    0.035  &    5250 $\pm $  200   \\ 
          \object{TrES-2}  &  2.47063  &  0.03670  &  1.221 $\pm $  0.044  &  1.198 $\pm $  0.053  &  1.000 $\pm $  0.035  &  0.980 $\pm $    0.062  &    5850 $\pm $   50   \\ 
          \object{TrES-3}  &  1.30619  &  0.02280  &  1.336 $\pm $  0.034  &  1.910 $\pm $  0.080  &  0.830 $\pm $  0.020  &  0.930 $\pm $    0.040  &    5650 $\pm $   75   \\ 
          \object{TrES-4}  &  3.55395  &  0.05100  &  1.783 $\pm $  0.090  &  0.925 $\pm $  0.080  &  1.849 $\pm $  0.090  &  1.400 $\pm $    0.100  &    6200 $\pm $   75   \\ 
            \object{XO-1}  &  3.94153  &  0.04880  &  1.189 $\pm $  0.023  &  0.900 $\pm $  0.070  &  0.928 $\pm $  0.033  &  1.000 $\pm $    0.030  &    5750 $\pm $   13   \\ 
            \object{XO-2}  &  2.61584  &  0.03680  &  0.984 $\pm $  0.019  &  0.570 $\pm $  0.060  &  0.964 $\pm $  0.020  &  0.980 $\pm $    0.030  &    5340 $\pm $   32   \\ 
            \object{XO-4}  &  4.12502  &  0.05550  &  1.340 $\pm $  0.050  &  1.720 $\pm $  0.200  &  1.560 $\pm $  0.050  &  1.320 $\pm $    0.020  &    6397 $\pm $   70   \\ 
            \object{XO-5}  &  4.18775  &  0.04880  &  1.089 $\pm $  0.057  &  1.077 $\pm $  0.037  &  1.080 $\pm $  0.050  &  0.880 $\pm $    0.030  &    5370 $\pm $   70   \\ 
         \object{HAT-P-1}  &  4.46529  &  0.05510  &  1.203 $\pm $  0.051  &  0.530 $\pm $  0.040  &  1.115 $\pm $  0.043  &  1.120 $\pm $    0.090  &    5975 $\pm $   45   \\ 
         \object{HAT-P-3}  &  2.89970  &  0.03894  &  0.890 $\pm $  0.046  &  0.599 $\pm $  0.026  &  0.824 $\pm $  0.040  &  0.936 $\pm $    0.050  &    5185 $\pm $   46   \\ 
         \object{HAT-P-4}  &  3.05654  &  0.04460  &  1.270 $\pm $  0.050  &  0.680 $\pm $  0.040  &  1.590 $\pm $  0.070  &  1.260 $\pm $    0.100  &    5860 $\pm $   80   \\ 
         \object{HAT-P-5}  &  2.78849  &  0.04075  &  1.260 $\pm $  0.050  &  1.060 $\pm $  0.110  &  1.170 $\pm $  0.050  &  1.160 $\pm $    0.060  &    5960 $\pm $  100   \\ 
         \object{HAT-P-6}  &  3.85298  &  0.05235  &  1.330 $\pm $  0.061  &  1.057 $\pm $  0.119  &  1.460 $\pm $  0.060  &  1.290 $\pm $    0.060  &    6570 $\pm $   80   \\ 
         \object{HAT-P-8}  &  3.07632  &  0.04870  &  1.460 $\pm $  0.020  &  1.520 $\pm $  0.170  &  1.580 $\pm $  0.070  &  1.280 $\pm $    0.040  &    6200 $\pm $   80   \\ 
         \object{HAT-P-9}  &  3.92289  &  0.05300  &  1.400 $\pm $  0.060  &  0.780 $\pm $  0.090  &  1.320 $\pm $  0.070  &  1.280 $\pm $    0.130  &    6350 $\pm $  150   \\ 
  \object{HAT-P-10/WASP11}  &  3.72247  &  0.04350  &  1.055 $\pm $  0.040  &  0.487 $\pm $  0.030  &  0.790 $\pm $  0.030  &  0.830 $\pm $    0.030  &    4980 $\pm $   60   \\ 
        \object{HAT-P-12}  &  3.21306  &  0.03840  &  0.963 $\pm $  0.025  &  0.211 $\pm $  0.012  &  0.700 $\pm $  0.030  &  0.730 $\pm $    0.020  &    4650 $\pm $   60   \\ 
        \object{HAT-P-13}  &  2.91626  &  0.04260  &  1.280 $\pm $  0.080  &  0.851 $\pm $  0.046  &  1.560 $\pm $  0.080  &  1.220 $\pm $    0.100  &    5638 $\pm $   90   \\ 
          \object{WASP-1}  &  2.51996  &  0.03820  &  1.443 $\pm $  0.039  &  0.790 $\pm $  0.130  &  1.453 $\pm $  0.032  &  1.150 $\pm $    0.090  &    6110 $\pm $  245   \\ 
          \object{WASP-2}  &  2.15223  &  0.03070  &  1.038 $\pm $  0.050  &  0.880 $\pm $  0.070  &  0.780 $\pm $  0.060  &  0.790 $\pm $    0.150  &    5200 $\pm $  200   \\ 
          \object{WASP-3}  &  1.84684  &  0.03170  &  1.255 $\pm $  0.085  &  1.760 $\pm $  0.090  &  1.310 $\pm $  0.090  &  1.240 $\pm $    0.090  &    6400 $\pm $  100   \\ 
          \object{WASP-4}  &  1.33823  &  0.02255  &  1.602 $\pm $  0.258  &  1.237 $\pm $  0.064  &  0.873 $\pm $  0.036  &  0.850 $\pm $    0.120  &    5500 $\pm $  100   \\ 
          \object{WASP-5}  &  1.62843  &  0.02670  &  1.100 $\pm $  0.070  &  1.580 $\pm $  0.110  &  1.030 $\pm $  0.060  &  0.970 $\pm $    0.090  &    5700 $\pm $  150   \\ 
          \object{WASP-6}  &  3.36101  &  0.04210  &  1.220 $\pm $  0.050  &  0.500 $\pm $  0.030  &  0.870 $\pm $  0.030  &  0.880 $\pm $    0.070  &    5450 $\pm $  100   \\ 
          \object{WASP-8}  &  8.15875  &  0.07930  &  1.170 $\pm $  0.100  &  2.230 $\pm $  0.100  &    &    &    5600 $\pm $  300   \\ 
         \object{WASP-10}  &  3.09276  &  0.03781  &  1.080 $\pm $  0.020  &  3.150 $\pm $  0.120  &  0.698 $\pm $  0.012  &  0.750 $\pm $    0.030  &    4675 $\pm $  100   \\ 
         \object{WASP-12}  &  1.09142  &  0.02290  &  1.790 $\pm $  0.090  &  1.410 $\pm $  0.090  &  1.570 $\pm $  0.070  &  1.350 $\pm $    0.140  &    6250 $\pm $  150   \\ 
         \object{WASP-13}  &  4.35298  &  0.05270  &  1.220 $\pm $  0.130  &  0.460 $\pm $  0.060  &  1.340 $\pm $  0.130  &  1.030 $\pm $    0.110  &    5826 $\pm $  100   \\ 
         \object{WASP-15}  &  3.75207  &  0.04990  &  1.430 $\pm $  0.080  &  0.540 $\pm $  0.050  &  1.477 $\pm $  0.070  &  1.180 $\pm $    0.120  &    6300 $\pm $  100   \\ 
         \object{WASP-16}  &  3.11806  &  0.04210  &  1.019 $\pm $  0.072  &  0.855 $\pm $  0.045  &  0.946 $\pm $  0.060  &  1.020 $\pm $    0.070  &    5700 $\pm $  150   \\ 
         \object{WASP-18}  &  0.94145  &  0.02026  &  1.115 $\pm $  0.063  & 10.300 $\pm $  0.690  &  1.216 $\pm $  0.070  &  1.250 $\pm $    0.130  &    6400 $\pm $  100   \\ 
         \object{CoRoT-2}  &  1.74300  &  0.02800  &  1.465 $\pm $  0.029  &  3.310 $\pm $  0.160  &  0.902 $\pm $  0.018  &  0.970 $\pm $    0.060  &    5625 $\pm $  120   \\ 
         \object{CoRoT-3}  &  4.25680  &  0.05694  &  1.060 $\pm $  0.120  & 21.660 $\pm $  1.000  &  1.540 $\pm $  0.090  &  1.360 $\pm $    0.090  &    6740 $\pm $  140   \\ 
         \object{CoRoT-4}  &  9.20205  &  0.09000  &  1.190 $\pm $  0.060  &  0.720 $\pm $  0.080  &  1.170 $\pm $  0.020  &  1.160 $\pm $    0.030  &    6190 $\pm $   60   \\ 
         \object{CoRoT-5}  &  4.03790  &  0.04947  &  1.388 $\pm $  0.046  &  0.470 $\pm $  0.050  &  1.186 $\pm $  0.040  &  1.000 $\pm $    0.020  &    6100 $\pm $   65   \\ 
      \object{OGLE-TR-L9}  &  2.48553  &  0.03080  &  1.610 $\pm $  0.040  &  4.500 $\pm $  1.500  &  1.530 $\pm $  0.040  &  1.520 $\pm $    0.080  &    6933 $\pm $   60   \\ 
      \object{OGLE-TR-10}  &  3.10128  &  0.04162  &  1.245 $\pm $  0.095  &  0.610 $\pm $  0.130  &  1.140 $\pm $  0.080  &  1.100 $\pm $    0.050  &    6075 $\pm $   86   \\ 
      \object{OGLE-TR-56}  &  1.21191  &  0.02250  &  1.300 $\pm $  0.050  &  1.290 $\pm $  0.120  &  1.320 $\pm $  0.060  &  1.170 $\pm $    0.040  &    6119 $\pm $   62   \\ 
     \object{OGLE-TR-111}  &  4.01445  &  0.04670  &  1.010 $\pm $  0.040  &  0.520 $\pm $  0.130  &  0.831 $\pm $  0.031  &  0.810 $\pm $    0.020  &    5044 $\pm $   83   \\ 
     \object{OGLE-TR-113}  &  1.43248  &  0.02290  &  1.090 $\pm $  0.030  &  1.350 $\pm $  0.190  &  0.770 $\pm $  0.020  &  0.780 $\pm $    0.020  &    4804 $\pm $  106   \\ 
     \object{OGLE-TR-132}  &  1.68987  &  0.02990  &  1.180 $\pm $  0.070  &  1.140 $\pm $  0.120  &  1.340 $\pm $  0.080  &  1.260 $\pm $    0.030  &    6210 $\pm $   59   \\ 
     \object{OGLE-TR-182}  &  3.97910  &  0.05100  &  1.210 $\pm $  0.160  &  1.010 $\pm $  0.115  &  1.140 $\pm $  0.150  &  1.140 $\pm $    0.050  &    5924 $\pm $   64   \\ 
     \object{OGLE-TR-211}  &  3.67724  &  0.05100  &  1.395 $\pm $  0.125  &  1.030 $\pm $  0.200  &  1.640 $\pm $  0.130  &  1.330 $\pm $    0.050  &    6325 $\pm $   91   \\ 
   & & & & & & & \\ 
                           \end{tabular}
\caption{Parameters of the considered transiting planetary systems.}
\label{table_p1}
                             \end{table*}

                           \begin{table*}
                \begin{tabular}{lcccccr}
                         & & & & & & \\ 
                           Name & $ v \sin i$ & $P_{\rm rot}$ & Age & $\tau_{\rm synch}$ & $e$ & References \\
                                                               & (kms$^{-1}$) & (days) & (Gyr) & (Gyr) &  & \\
                         & & & & & & \\ 
        HD149026  &   6.00 $\pm $ 0.50  &   &   2.00- 3.60  &    34.10  &  0.0   &        Wo07 Sa05       \\ 
        HD189733  &            & 11.953 $\pm$ 0.009  &               &           &  0.004 $\pm $  0.003  &  He07 Mo07 Po07 Wi07   \\ 
   HD197286/WASP7  &  17.00 $\pm $ 2.00  &   &               &           &  0.0  &           He09         \\ 
        HD209458  &     & 11.4 $\pm$ 1.5  &               &           &  0.015 $\pm $  0.005  &       Wi05 Wi05a       \\ 
          TrES-1  &   1.30 $\pm $ 0.30  &   &               &           &  0.0   &    Na07 Wi07a Al04     \\ 
          TrES-2  &   1.00 $\pm $ 0.60  &   &   2.80- 7.80  &     0.56  &  0.0   &        Wi08 So07       \\ 
          TrES-3  &   1.50 $\pm $ 1.00  &   &   0.10- 3.70  &     0.04  &  0.0   &           So09         \\ 
          TrES-4  &   8.50 $\pm $ 0.50  &   &   2.50- 4.40  &    10.45  &  0.0  &           So09         \\ 
            XO-1  &   1.11 $\pm $ 0.70  &   &   0.60- 5.50  &     9.31  &  0.0   &        Ho06 Mc06       \\ 
            XO-2  &   1.30 $\pm $ 0.50  &   &               &           &  0.0   &        Fe09 Bu07       \\ 
            XO-4  &   8.80 $\pm $ 0.50  &   &   1.50- 2.70  &    12.03  &  0.0   &           Mc08         \\ 
            XO-5  &   0.70 $\pm $ 0.50  &   &  12.80-16.80  &     1.88  &  0.010 $\pm $  0.013  &           Pa09         \\ 
         HAT-P-1  &   3.75 $\pm $ 0.58  &   &   1.60- 4.60  &   130.20  &  0.0   &   Jo08 Wi07a Ba07a     \\ 
         HAT-P-3  &   0.50 $\pm $ 0.50  &   &   0.10- 6.90  &           &  0.0   &           To07         \\ 
         HAT-P-4  &   5.50 $\pm $ 0.50  &   &   3.60- 6.80  &     7.87  &  0.0   &        To08 Ko07       \\ 
         HAT-P-5  &   2.60 $\pm $ 1.50  &   &   0.80- 4.40  &     2.53  &  0.0   &       To08 Ba07b       \\ 
         HAT-P-6  &   8.70 $\pm $ 1.00  &   &   1.60- 2.80  &    28.17  &  0.0   &        To08 No08       \\ 
         HAT-P-8  &  11.50 $\pm $ 0.50  &   &   2.40- 4.40  &     7.33  &  0.0   &           La08         \\ 
         HAT-P-9  &  11.90 $\pm $ 1.00  &   &   0.20- 3.40  &   204.30  &  0.0   &        Am09 Sh09       \\ 
  HAT-P-10/WASP11  &   0.50 $\pm $ 0.20  &   &   4.10-11.70  &    11.05  &  0.0   &       We09 Ba09a       \\ 
        HAT-P-12  &   0.50 $\pm $ 0.40  &   &   0.50- 4.50  &    38.73  &  0.0   &           Ha09         \\ 
        HAT-P-13  &   4.10 $\pm $ 0.50  &   &   4.20- 7.50  &     2.79  &  0.021 $\pm $  0.009  &          Ba09b         \\ 
          WASP-1  &   5.79 $\pm $ 0.35  &   &   1.00- 3.00  &     3.11  &  0.0   &  Ca07 St07 Sh07 Ch07   \\ 
          WASP-2  &       &       &               &           &  0.0   &  Ca07 St07 Sh07 Ch07   \\ 
          WASP-3  &  13.40 $\pm $ 1.50  &   &               &           &  0.0   &        Gi08 Po08       \\ 
          WASP-4  &   2.00 $\pm $ 1.00  &   &               &           &  0.0   &           Gi09         \\ 
          WASP-5  &   3.50 $\pm $ 1.00  &   &               &           &  0.038 $\pm $  0.026  &           Gi09         \\ 
          WASP-6  &   1.60 $\pm $ 0.30  &   &               &           &  0.054 $\pm $  0.018  &          Gi09a         \\ 
          WASP-8  &       &       &               &           &  0.0   &           Sm09         \\ 
         WASP-10  &     & 11.91 $\pm$ 0.05  &               &           &  0.059 $\pm $  0.014  &        Jo09 Sm09       \\ 
         WASP-12  &   2.20 $\pm $ 1.50  &   &               &           &  0.049 $\pm $  0.015  &           He09         \\ 
         WASP-13  &   2.50 $\pm $ 2.50  &   &   3.60-14.00  &           &  0.0   &           Sk09         \\ 
         WASP-15  &   4.00 $\pm $ 2.00  &   &   2.60- 6.70  &    22.24  &  0.0   &           We09a         \\ 
         WASP-16  &   3.00 $\pm $ 1.00  &   &   0.10- 8.10  &    12.21  &  0.0   &           Li09         \\ 
         WASP-18  &  11.00 $\pm $ 1.50  &   &   0.50- 1.50  &  $6.5\times 10^{-4}$  &  0.009 $\pm $  0.003  &          He09a         \\ 
         CoRoT-2  &    & 4.52 $\pm$ 0.15  &   0.10- 1.70  &     0.36  &  0.0   &        Bo08 Al08       \\ 
         CoRoT-3  &  17.00 $\pm $ 1.00  &   &   1.60- 2.80  &     1.41  &  0.008 $\pm $  0.015  &        De08 Tr09       \\ 
         CoRoT-4  &    & 9.20 $\pm$ 0.3  &   0.70- 2.00  &  2728.00  &  0.0 $\pm $  0.100  &     Ai08 Mo08 La09     \\ 
         CoRoT-5  &   1.00 $\pm $ 1.00  &   &   5.50- 8.50  &           &  0.090 $\pm $  0.090  &           Ra09         \\ 
      OGLE-TR-L9  &  39.33 $\pm $ 0.40  &   &   0.10- 0.70  &     0.92  &  0.0   &           Sn09         \\ 
      OGLE-TR-10  &   7.70 $\pm $ 3.00  &   &               &           &  0.0   & To04 Bo05 Ko05 Po07a   \\ 
      OGLE-TR-56  &   3.00 $\pm $ 3.00  &   &               &           &  0.0   & To04 Bo05 Ko05 Po07a   \\ 
     OGLE-TR-111  &       &       &               &           &  0.0   &        Po04 Mi07       \\ 
     OGLE-TR-113  &   3.00 $\pm $ 3.00  &   &               &           &  0.0  &     Bo04 Di07 Ko04     \\ 
     OGLE-TR-132  &   3.00 $\pm $ 3.00  &   &   0.10- 1.40  &           &  0.0   &     Gi07 Mo04 Bo04     \\ 
     OGLE-TR-182  &       &       &               &           &  0.0   &          Po08a         \\ 
     OGLE-TR-211  &       &       &               &           &  0.0   &           Ud08         \\ 
                           \end{tabular}
\caption{Parameters of the considered transiting planetary systems.}
\begin{tiny}
    Reference codes: 
   Ai08:          \citet{Aigrainetal08}; 
   Al04:           \citet{Alonsoetal04}; 
   Al08:           \citet{Alonsoetal08}; 
   Am09:        \citet{Ammler-vonetal09};
   Ba07:         \citet{Bakosetal07};  
  Ba07a:           \citet{Bakosetal07a}; 
  Ba07b:           \citet{Bakosetal07b}; 
  Ba09a:           \citet{Bakosetal09a}; 
  Ba09b:           \citet{Bakosetal09c}; 
   Bo04:           \citet{Bouchyetal04}; 
   Bo05:           \citet{Bouchyetal05}; 
   Bo08:           \citet{Bouchyetal08}; 
   Bu07:            \citet{Burkeetal07}; 
   Ca07:          \citet{Cameronetal07}; 
   Ch07:      \citet{Charbonneauetal07}; 
   De08:          \citet{Deleuiletal08}; 
   Di07:             \citet{Diazetal07}; 
   Fe09:        \citet{Fernandezetal09}; 
   Gi07:           \citet{Gillonetal07}; 
   Gi08:           \citet{Gibsonetal08}; 
   Gi09:           \citet{Gillonetal09}; 
  Gi09a:          \citet{Gillonetal09a}; 
   Ha09:          \citet{Hartmanetal09}; 
   He08:            \citet{HenryWinn08}; 
   He09:             \citet{Hebbetal09}; 
  He09a:          \citet{Hellieretal09}; 
  He09b:         \citet{Hellieretal09a}; 
   Ho06:           \citet{Holmanetal06}; 
   Jo08:          \citet{Johnsonetal08}; 
   Jo09:            \citet{Joshietal09}; 
   Ko04:          \citet{Konackietal04}; 
   Ko05:          \citet{Konackietal05}; 
   Ko07:           \citet{Kovacsetal07}; 
   La08:           \citet{Lathametal08}; 
   La09:            \citet{Lanzaetal09b}; 
   Li09:           \citet{Listeretal09}; 
   Mc06:       \citet{McCulloughetal06}; 
   Mc08:       \citet{McCulloughetal08}; 
   Mi07:          \citet{Minnitietal07}; 
   Mo04:           \citet{Moutouetal04}; 
   Mo07:           \citet{Moutouetal07}; 
   Mo08:           \citet{Moutouetal08}; 
   Na07:           \citet{Naritaetal07}; 
   No08:            \citet{Noyesetal08}; 
   Pa09:              \citet{Paletal09}; 
   Po04:             \citet{Pontetal04}; 
   Po07:             \citet{Pontetal07}; 
  Po07a:            \citet{Pontetal07a}; 
   Po08:         \citet{Pollaccoetal08}; 
  Po08a:            \citet{Pontetal08a}; 
   Ra09:            \citet{Raueretal09}; 
   Sa05:             \citet{Satoetal05}; 
   Sh07:          \citet{Shporeretal07}; 
   Sh09:          \citet{Shporeretal09}; 
   Sk09:          \citet{Skillenetal09}; 
   Sm09:            \citet{Smithetal09}; 
   Sn09:          \citet{Snellenetal09}; 
   So07:         \citet{Sozzettietal07}; 
   So09:         \citet{Sozzettietal09}; 
   St07:         \citet{Stempelsetal07}; 
   To04:           \citet{Torresetal04}; 
   To07:           \citet{Torresetal07}; 
   To08:           \citet{Torresetal08}; 
   Tr09:           \citet{Triaudetal09}; 
   Ud08:          \citet{Udalskietal08}; 
   We09:            \citet{Westetal09a}; 
   We09a:           \citet{Westetal09b}; 
   Wi05:             \citet{Winnetal05}; 
  Wi05a:       \citet{Wittenmyeretal05}; 
   Wi07:             \citet{Winnetal07}; 
  Wi07a:            \citet{Winnetal07a}; 
  Wi07b:            \citet{Winnetal07b}; 
   Wi08:             \citet{Winnetal08}; 
   Wo07:             \citet{Wolfetal07}. 
\end{tiny}
\label{table_p2}
                             \end{table*}

                           \begin{table}
               \begin{tabular}{ccccc}
                       & & & &  \\ 
              $(n/\Omega)_{\rm max}$ & $P_{1}$ & $N_{1}$ & $P_{2}$ & $N_{2}$ \\
                  & & & & \\ 
              4 & $3.30 \times 10^{-6}$ & 12 &  $-$  & 3 \\
              5 & $6.97 \times 10^{-6}$ & 14 & $-$ & 4 \\
              6 & $2.50 \times 10^{-6}$ & 16 & 0.155 & 6 \\
              7 & $2.50 \times 10^{-6}$ & 16 & 0.229 & 7 \\
              10 & $2.50 \times 10^{-6}$ & 16 & 0.450 & 10 \\
              12 & $2.50 \times 10^{-6}$ & 16 & 0.575 & 12 \\
                 & & & & \\ 
                           \end{tabular}
\caption{Kolmogorov-Smirnov test of uniform distribution for different subsamples of the distribution of $n/\Omega$.}
\label{KS_test}
                             \end{table}

\section{Modelling the angular momentum content of a stellar corona}

\subsection{Coronal field model}
\label{field_model}

We adopt  a spherical polar coordinate frame having its  origin at the barycentre of the host star and the polar axis along the stellar rotation axis. The radial distance from the origin is indicated with $r$,  the colatitude measured from the North pole with $\theta$, and the azimuthal angle with $\phi$. The planet orbit is assumed circular  and lying in the equatorial plane of the star according to the selection criteria considered in Sect.~\ref{observations}. We adopt a reference frame rotating with the angular velocity of the star  $\Omega$ with respect to an inertial frame. 

{ We model the coronal field under the hypothesis that} the magnetic pressure  is much greater than the plasma pressure and the gravitational force, so we can assume  a force-free magnetohydrostatic balance, i.e., the current density ${\vec J}$ is everywhere parallel to the magnetic field ${\vec B}$, viz. ${\vec J} \times {\vec B} = 0$. This means that $\nabla \times {\vec B} = \alpha {\vec B}$, with  the force-free parameter $\alpha$ constant along each field line \citep{Priest82}. If $\alpha$ is uniform in the stellar corona, the field is called a linear force-free field and it satisfies the vector Helmoltz equation $\nabla^{2} {\vec B} + \alpha^{2} {\vec B} = 0$. Its solutions in spherical geometry have been studied by, e.g.,  \citet{Chandrasekhar56} and \citet{ChandrasekharKendall57}. 

Linear force-free fields are particularly attractive in view of their mathematical symplicity and their minimum-energy properties in a finite domain that contains all the lines of force, as shown by \citet{Woltjer58}. Specifically, in ideal magnetohydrodynamics, the minimum energy state of a magnetic field in a finite domain is a linear force-free state set according to the boundary conditions and the  conservation of magnetic helicity. \citet{Berger85} modified the definition of magnetic helicity introducing a relative magnetic helicity which is conserved in spite of the fact that the lines of force may not be contained into a finite volume. Such a relative helicity is the relevant conserved quantity in the case of a stellar corona whose lines of force cross the surface of the star. 

Considering the conservation of relative helicity, e.g., \citet{HeyvaertsPriest84} and \citet{RegnierPriest07} have conjectured that the minimum energy state actually allowable to a solar active region in a semi-infinite atmosphere is a linear force-free state \citep[see, however, discussion in ][]{ZhangLow05}. 
We conjecture that the enhanced dissipation induced by the motion of a hot Jupiter inside a stellar corona drives the coronal field toward such a linear force-free state \citep[cf. ][]{Lanza09}. 
{ Note that, even in the case of a synchronous system such as $\tau$~Boo or CoRoT-4, the differential rotation of the  stellar surface implies relative velocities of the order of $10-30$ km s$^{-1}$ between the planet and the coronal field \citep[cf., ][]{Catalaetal07,Lanzaetal09b}. 
}  

To model the  stellar coronal field, we  consider only the dipole-like component (i.e., with a radial order $k=1$) of the linear force-free solution of \citet{ChandrasekharKendall57} because it has the slowest decay with distance from the star and therefore contains most of the angular momentum of the corona (see below and Sect.~\ref{angular_mom}). 
 Moreover,  an axisymmetric field (i.e., with an azimuthal degree $m=0$) is the simplest geometry to model the corona and is also favoured by models of magnetic star-planet interaction, as discussed by \citet{Lanza08,Lanza09}.
{ 
In particular, an axisymmetric field reproduces the observed phase lag between the planet and the chromospheric hot spot induced by its interaction with the coronal field of the star \citep[e.g., ][]{Shkolniketal05,Shkolniketal08}.
}  

Our linear force-free field can be expressed in the formulism of \citet{Flyeretal04} as: 
\begin{equation}
{\vec B} = \frac{1}{r \sin \theta} \left[ \frac{1}{r} \frac{\partial A}{\partial \theta} \hat{\vec r} - \frac{\partial A}{\partial r} \hat{\vec \theta} + Q(A) \hat{\vec \phi} \right],
\label{field_express}
\end{equation}
where $A(r, \theta)$ is the flux function of the field and $Q=\alpha A$. 
Magnetic field lines lie  over surfaces of constant $A(r, \theta)$, 
as follows by noting that ${\vec B} \cdot \nabla A = 0$. The flux function for our dipole-like field geometry is  $A(r, \theta) = B_{0} R^{2} g(q) \sin^{2} \theta$, where
$2 B_{0}$ is the magnetic field intensity at the North pole of the star, $R$ the star's radius and the function $g(q)$ is defined by:
\begin{equation}
g (q) \equiv \frac{[b_{0} J_{-3/2}(q) + c_{0} J_{3/2}(q)]\sqrt{q}}{[b_{0} J_{-3/2}(q_{0}) + c_{0} J_{3/2}(q_{0})] \sqrt{q_{0}}},
\end{equation}
where $b_{0}$ and $c_{0}$ are free coefficients, $J_{-3/2}$ and $J_{3/2}$ are Bessel functions of the first kind of order $-3/2$ and $3/2$, respectively, $q \equiv |\alpha | r$, and $q_{0} \equiv |\alpha | R$. Making use of Eq.~(\ref{field_express}), the magnetic field components are:
\begin{eqnarray}
B_{r} & = & 2B_{0} \frac{R^{2}}{r^{2}} g(q) \cos \theta, \nonumber \\
\label{field_conf}
B_{\theta}  & = & -B_{0} |\alpha | \frac{R^{2}}{r} g^{\prime} (q)  \sin \theta, \\
B_{\phi} & = & \alpha B_{0} \frac{R^{2}}{r} g(q)  \sin \theta \nonumber,  
\end{eqnarray}
where $g^{\prime} (q) \equiv dg/dq$. 
A linear force-free field as given by Eqs.~(\ref{field_conf}) extends to the infinity with an infinity energy. We consider its restriction to the radial domain $q_{0} \leq q \leq q_{\rm L}$, where $q_{\rm L}$ is the first zero of $g(q)$, so that all the  magnetic field lines are closed  in our model \citep[see ][ for the boundary conditions at $r=r_{\rm L} \equiv q_{\rm L}/ |\alpha | $]{Chandrasekhar56}. 

The magnetic field geometry specified by Eqs.~(\ref{field_conf}) depends on two independent parameters, i.e., $\alpha$ and $b_{0}/c_{0}$. They can be derived from the boundary conditions at the stellar photosphere, i.e., knowing the magnetic field ${\vec B}^{(s)}(\theta, \phi) $ on the surface at $r=R$. Using the orthogonality properties of the basic poloidal and toroidal fields \citep[see ][]{Chandrasekhar61}, we find: 
\begin{eqnarray}
\frac{ 8 \pi}{3} B_{0} R^{2} &  = & \int_{\Sigma(R)} B_{\rm r}^{(s)} \cos \theta d \Sigma,  \nonumber \\
\label{BC_eqs}
  \frac{ 8 \pi}{3} |\alpha | B_{0} R^{3} g^{\prime} (q_{0}) & = & - \int_{\Sigma(R)} B_{\theta}^{(s)} \sin \theta d \Sigma,  \\
\frac{ 8 \pi}{3} \alpha  B_{0} R^{3} & = & - \int_{\Sigma(R)} B_{\phi}^{(s)} \sin \theta d \Sigma, \nonumber 
\end{eqnarray}  
where  $\Sigma (R)$ is the spherical surface of radius $R$, and $d \Sigma = R^{2} \sin \theta d\theta d \phi$. Note that the photospheric magnetic field components can be measured by means of spectropolarimetric techniques if the star rotates fast enough ($v \sin i \geq 10-15$~km~s$^{-1}$) as shown in the case of, e.g., $\tau$ Boo by \citet{Catalaetal07} and \citet{Donatietal08}. { With such a kind of observations}, Eqs.~(\ref{BC_eqs}) can be applied to derive the parameters of the coronal field model and its topology, provided that the field is approximately force-free down to the stellar photosphere. { We shall refer to this approach in Sect.~\ref{ms_evol}, where our model will be used to study the evolution of stellar angular momentum.}

The magnetic energy $E$ of the field confined between the spherical surfaces $r=R$ and $r=r_{\rm L}$ can be found from Eq.~(79) in \S~40 of \citet{Chandrasekhar61}:
\begin{equation}
E = E_{\rm p} \left\{ 2 + q_{0} q_{\rm L} [g^{\prime}(q_{\rm L})]^{2} - q_{0}^{2} [g^{\prime}(q_{0})]^{2} - q_{0}^{2} \right\}, 
\label{Benergy}
\end{equation}
where $E_{\rm p} \equiv ({4 \pi}/{3 \mu}) R^{3} B_{0}^{2}$ is the energy of the potential dipole field with the same radial component at the surface $r=R$, and $\mu$ is the magnetic permeability. The relative magnetic helicity $H_{\rm R}$, as defined by \citet{Berger85}, can be found from his Eq.~(19) and is:
\begin{equation}
H_{\rm R} = B_{0}^{2} R^{4} \left[ 2 g^{\prime}(q_{0}) + \frac{8 \pi}{3} \frac{E}{q_{0} E_{\rm p}} \right] \frac{|\alpha|}{\alpha}. 
\label{Bhelic}
\end{equation} 
Note that the field obtained by changing the sign of $\alpha$ has the same poloidal components $B_{\rm r}$ and $B_{\theta}$, and energy $E$, while the toroidal component $B_{\phi}$ and the relative helicity $H_{\rm R}$ become opposite. Further information on the field described by 
Eqs.~(\ref{field_conf})  can be found in, e.g., \citet[][]{Lanza09}. 

For a finite $\alpha$, $E > E_{\rm p}$ because the potential field has the minimum energy for a given $B_{\rm r}^{(s)}$. If we consider all magnetic fields with one end of their field lines  anchored at $r=R$ and the other out to the infinity, satisfying the same boundary conditions of our field at $r=R$, the field with the lowest possible energy is called the Aly field and its energy $E_{\rm Aly} = 1.66 E_{\rm p}$ \citep[see ][]{Flyeretal04}. We assume that the Aly energy is an upper bound for the energy of our field because it is the lowest energy allowing the field to open up all its lines of force out to the infinity driving a plasma outflow similar to a solar coronal mass ejection.  

Since all magnetic field lines are closed in our model, it is not possible to have a steady flux of angular momentum toward the infinity as, e.g., in the open field configuration of solar coronal holes. Therefore, we assume that the loss of angular momentum occurs only when the field energy reaches the Aly energy and the coronal field opens up toward the infinity driving out all the coronal mass. The rate of angular momentum loss depends on the angular momentum stored in the coronal field and the rate of occurrence of such events that we call coronal mass ejections (CMEs) by analogy with similar solar events which usually involve only a single active region and not the whole coronal field as in our simplified model. 

\begin{figure}[t]
\includegraphics[width=8cm,height=8cm]{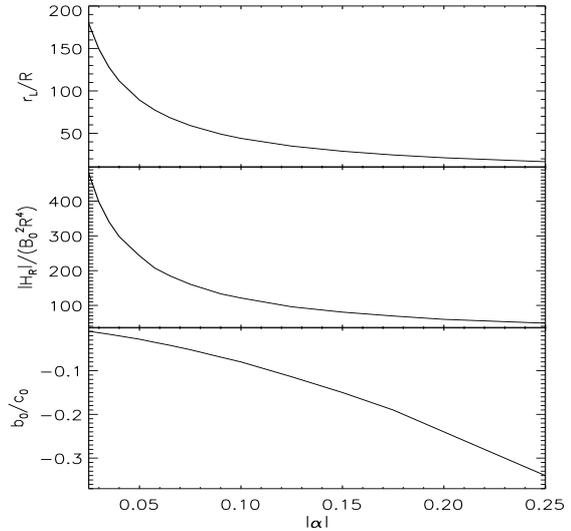}
\caption{Upper panel: The outer radius $r_{\rm L}$ of the coronal field at the Aly energy limit vs. the absolute value of the force-free
parameter $\alpha$; middle panel: the absolute value of the relative magnetic helicity  of the field at the Aly energy limit $|H_{\rm R}|$,  as computed 
from Eq.~(\ref{Bhelic}), vs. $| \alpha |$; lower panel: the value of the parameter $b_{0}/c_{0}$ corresponding to the Aly energy limit vs. $| \alpha |$.}
\label{rmax}
\end{figure}

If we fix the value of the parameter $\alpha$, the value of the ratio $b_{0}/c_{0}$ corresponding to the Aly energy can be determined numerically. We plot in Fig.~\ref{rmax} the outer radius, the relative magnetic helicity and the value of 
$b_{0}/c_{0}$ vs. $\alpha$ for a field at the Aly energy limit. There is a remarkable decrease of the outer field radius and of the relative magnetic helicity with increasing $\alpha$. 
This implies that a coronal configuration with a greater $\alpha$ is more tightly confined than one with a lower value of $\alpha$. Moreover, its relative helicity  is significantly smaller than in the case with a lower $\alpha$.

\subsection{Some considerations on non-linear force-free field models}

The application of our linear model is justified in view of its  mathematical simplicity and the hypothesis of a corona with closed field lines. A  general treatment of the coronal field topology poses formidable mathematical problems even considering only force-free fields.  Therefore, we restrict ourselves to a special class of non-linear force-free models to make some progress. 
 
\citet{LowLou00}, \citet{Flyeretal04}, \citet{Zhangetal06}, and other authors, showed that a force-free field can extend to the infinity with a finite magnetic energy and relative helicity if $\alpha$ is not uniform, i.e., it is a non-linear force-free field. Such fields are suitable to model a stellar corona without the limitations of our adopted linear force-free model, specifically the property that all field lines are closed. An example of such a non-linear axisymmetric field has been provided 
by  \citet{LowLou00}. It is obtained by assuming $Q(A)=\Theta A^{1 + \frac{1}{n}}$ in Eq.~(\ref{field_express}), with $\Theta$ being a constant and $A(r, \theta) = P(\theta) r^{-n}$; i.e., the flux function $A$ is of separable form, with $P$ being a continuous function of $\theta$ defined in the closed interval $[0, \pi]$ which must vanish at the poles to ensure the regularity of the field.  

Let us consider a magnetic field line going from the surface of the star to the infinity.  Since $A$ must be constant on it, we conclude that $P$ must be zero, otherwise $P$ would increase without bound following the field line for $n >0$, which is of course not possible in view of its continuity.  In other words, all the magnetic field lines going out to the  infinity must be rooted on the surface of the star at colatitudes where $P(\theta)=0$. 

Isolated field lines going out to the infinity do not have any  effect on the stellar angular momentum loss because the mass flow along them vanishes. To produce an angular momentum loss, we need a flux tube with a finite cross-section, i.e., a bundle of field lines covering a finite colatitude interval on the surface of the star, say, $\theta_{1} \leq \theta \leq \theta_{2}$, in which $P(\theta) =0$. Considering the expression for $B_{\rm r}$ in Eq.~(\ref{field_express}), it follows that $B_{\rm r} = 0$ inside that interval because $(\partial A/\partial \theta)_{r=R} = 0$. In other words, those field lines do not intersect the surface of the star which is the source of any wind mass loss. We deduce that there can be no steady mass loss along those field lines, hence their contribution to the angular momentum loss is negligible. 

In conclusion, a non-linear force-free field with a separable flux function of the kind proposed by \citet{LowLou00}, although endowed with field lines going from the surface of the star to the infinity, cannot sustain a steady angular momentum loss from the star through a continuous wind flow. The only available mechanism is therefore represented by  the CME events considered above. 

Of course such a conclusion has been obtained for a specific class of non-linear models and cannot be generalized to all possible non-linear fields. Nevertheless, it shows that there are non-linear force-free fields for which our approach of considering only the angular momentum stored into a closed-field corona is  valid. Our linear model has the advantage of a spherically symmetric outer boundary of the corona at $r=r_{\rm L}$, while non-linear models may have a boundary that depends on the colatitude.

\subsection{Angular momentum content of the coronal field}
\label{angular_mom}

To compute the angular momentum content of the plasma trapped into the closed coronal 
field introduced in Sect.~\ref{field_model}, we need to evaluate its moment of inertia. 
For the sake of simplicity, let us assume that the plasma has a uniform temperature $T$ and is in hydrostatic equilibrium along each magnetic field line. The equation of hydrostatic equilibrium  reads \citep[cf., e.g., ][]{Priest82}:
\begin{equation}
\hat{\vec s}\cdot (- \nabla p + \rho \nabla \Phi) = 0,
\label{hydro_eq}
\end{equation}
where $\hat{\vec s}$ is the unit vector in the direction of the magnetic field, $p$ the plasma pressure, $\rho$ its density and $\Phi$ the total  potential including the gravitational and centrifugal terms. Since 
$p= ({\tilde{R}}/{\tilde{\mu}}) \rho T$ and $\Phi = G M /r + \frac{1}{2} \Omega r^{2} \sin^{2}  \theta$, where $\tilde{R}$ is the gas constant, $\tilde{\mu} $ the molecular weight of the plasma, $G$ the gravitation constant, and 
$M$ the mass of the star, Eq.~(\ref{hydro_eq}) can be immediately integrated  to give:
\begin{equation}
\rho = \rho_{0} \exp \left[ \frac{\tilde{\mu}}{\tilde{R} T} ( \Phi - \Phi_{0} ) \right],
\end{equation}
where $\rho_{0}$ and $\Phi_{0}$ are the density and the potential on a field line at the base of the corona, i.e., at a radius $r=r_{0}$. The density is maximum on the equatorial plane of the star, i.e., at $\theta= \pi/2$ because there $\Phi$ is maximum. For our purposes, we can estimate the maximum moment of inertia, which gives us an upper limit for the angular momentum loss, by assuming a spherically symmetric distribution of the density computed by evaluating $\Phi$ on the equatorial plane, i.e.: 
\begin{equation}
\rho(r) = \rho_{0} \exp \left\{ \frac{r_{0}}{H_{\rm p}} \left[ \left( \frac{r_{0}}{r} -1 \right) + \epsilon \left( \frac{r^{2}}{r_{0}^{2}} - 1 \right) \right] \right\}, 
\label{dens}
\end{equation}
where  
$H_{\rm p} \equiv \tilde{R} T/(\tilde{\mu} g_{0}) $ is the pressure scale height at the base of the corona, with $g=GM/r_{0}^{2}$ being the gravitational acceleration at $r=r_{0}$, and $\epsilon \equiv r_{0}^{3} \Omega^{2} /(2 G M)$  the ratio of the centrifugal to the gravitational potential at the base of the corona. 
We assume that the base of the corona is at $r_{0}=2R$, outward of which the plasma temperature is assumed to be constant.
Considering a star with the radius and mass of the Sun and a mean molecular weight $\tilde{\mu}=0.6$, we have  $H_{\rm p} \simeq 2.04 \times 10^{8} (T / 10^{6})$ m, with $T$ in K. As a typical coronal density, we adopt the mean electron density of the solar corona at $2R_{\odot}$, i.e., $n_{\rm e}=10^{12}$ m$^{-3}$ \citep{Allen00}.

The moment of inertia of the coronal plasma can be easily computed noting that in our assumptions the density is a function of $r$ only. If the corona extends up to the limit radius $r_{\rm L}$, its moment of inertia is:
\begin{equation}
I = \int_{V} \rho r^{2} \sin^{2} \theta \, dV, 
\label{mom_inertia}
\end{equation}
where $V$ is the volume of the corona, that is the spherical shell between radii $r_{0}$ and $r_{\rm L}$. 

\begin{figure}[t]
\includegraphics[width=8cm,height=8cm]{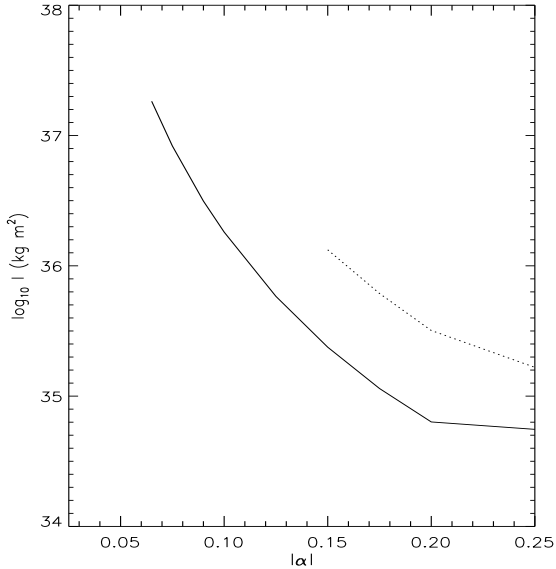}
\caption{The moment of inertia of the coronal plasma  vs. the absolute value of the force-free parameter $\alpha$, computed according to Eqs.~(\ref{dens}) and~(\ref{mom_inertia}) for a star with
$R=R_{\odot}$, $M=M_{\odot}$, $r_{0}= 2 R_{\odot}$, a base electron density $n_{\rm e} (r_{0}) =10^{12}$ m$^{-3}$, and temperature
$T= 1.6 \times 10^{6}$ K (solid line) or $T=3 \times 10^{6}$ K (dotted line).  The linear force-free configuration of the coronal field is that corresponding to the Aly energy limit for the given value of $\alpha$. The slope changes at $| \alpha | = 0.2 $ are due to the fixed value of $P_{\rm rot}=3$ days for $| \alpha | \geq 0.2$.  }
\label{mom_i}
\end{figure}

In Fig.~\ref{mom_i}, we plot the moment of inertia of a corona at the Aly energy limit vs. the force-free parameter $\alpha$ for a star analogous to the Sun, setting the coronal base at $r_{0}=2R$ with  $n_{\rm e}=10^{12}$ m$^{-3}$. We compute $I$ for two temperature values, i.e., $T=1.6 \times 10^{6}$ K, typical of stars with a low level of coronal emission as the Sun \citep[cf. ][]{Allen00}, and $T=3 \times 10^{6}$ K, which is characteristic of stars with a moderately high level of coronal emission, i.e., with an X-ray flux about one order of magnitude greater than the Sun at the maximum of the 11-yr cycle \citep[cf., e.g., ][]{Schmitt97}. We assume that the rotation period $P_{\rm rot}$ is inversely correlated with $| \alpha | $, as will be discussed in Sect.~\ref{ms_evol}.  Specifically, we assume that $P_{\rm rot}$ increases linearly between $3$ and $24$ days when $| \alpha |$ decreases from 0.2 to 0.025. 

The plots in Fig.~\ref{mom_i} are terminated where the potential energy or the internal energy of the plasma
exceed 0.1 of the total magnetic energy of the field computed for $B_{0} = 20$ G because the force-free condition is no longer valid in such a case. The potential energy $E_{\rm GC}$ and the internal energy $U$ of the coronal plasma are evaluated as:
\begin{equation}
E_{\rm GC} = \int_{V} \rho \Phi \, dV,
\end{equation} 
and
\begin{equation}
U = \frac{1}{\tilde{\gamma} - 1} \frac{\tilde{R}T}{\tilde{\mu}} \int_{V}  \rho \, dV,
\end{equation}
where $\tilde{\gamma}=5/3$ is the ratio of the specific heats of the plasma.  

Note the decrease of the moment of inertia by $\sim 2.5$ orders of magnitude when  
$| \alpha | $ increases from $0.06$ to $0.2$ owing to a remarkable decrease of the outer radius $r_{\rm L}$ of the corona (cf. Fig.~\ref{rmax}, upper panel). On the other hand, an increase of the  coronal temperature by a factor of $\sim 2$, as expected for rapidly rotating stars, produces an increase of the moment of inertia only by a factor of $\sim 3-4$.  The moment of inertia is directly proportial to the base density, so a change of $\rho_{0}$ by, say, one order of magnitude produces a corresponding change in the moment of inertia. We conclude that the most relevant variation of the moment of inertia of the corona is produced by a variation of the force-free parameter $\alpha$. 

\section{Application to stellar angular momentum evolution}
\label{application}

\subsection{Pre-main-sequence evolution}
\label{pms_evolution}
To apply the results of Sect.~\ref{angular_mom} to the problem of stellar angular momentum evolution, we need to define the initial rotation state of a star. { Stars with hot Jupiters are accompanied by  circumstellar discs during the pre-main-sequence (hereinafter PMS) phase of their evolution, which play a fundamental role in the formation and orbital evolution of their planets}. The angular velocity of a PMS star is equal to the Keplerian angular velocity 
 of its disc  at the so-called corotation radius. It is located $\sim 5-10$ percent outside the inner boundary of the disc, where it is truncated by the stellar magnetic field \citep[cf., e.g., ][ and references therein]{CameronCampbell93,Tinkeretal02,Scholzetal07,Bouvier08}.
According to the current theoretical scenario, hot Jupiters are formed at several AUs from their stars, beyond the snow line where volatile elements can condense, and then  migrate toward their stars on a timescale not exceeding $10^{5}-10^{6}$ yr { \citep[cf., e.g., ][]{PapaloizouTerquem06}}. If the stellar magnetic field is strong enough, the Keplerian shear induces a sizeable toroidal magnetic field in the ionized region of the disc close to the star and the inward migration of the planet may be halted close to the corotation radius, as suggested by \citet{Terquem03}. 
{ 
The field intensity required to halt inward migration depends on the variation of the  parameter $\beta $ vs. the radius within the disc, where $\beta=(c_{\rm s}/ v_{\rm A})^{2}$ with $c_{\rm s}$ being the sound speed and $v_{\rm A}$ the Alfv\'en speed. \citet{Terquem03} showed that values of $\beta \approx 10-100$ can be sufficient to halt planetary migration. 
}

\begin{figure}[t]
\includegraphics[width=8cm,height=8cm]{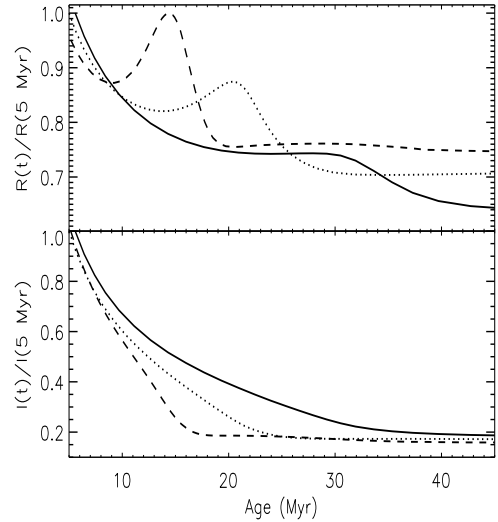}
\caption{Upper panel: Radius of PMS stars of different mass vs. time measured from their birth line; the radius is normalized to the value at an age of 5 Myr, corresponding to the average lifetime of the circumstellar discs; different linestyles refer to different masses: solid: $1$~M$_{\odot}$; dotted: 1.2~M$_{\odot}$; dashed: 
1.4~M$_{\odot}$. Lower panel: Moment of inertia of PMS stars vs. time from their birth; the moment of inertia is normalized at the value at an age of 5 Myr; different linestyles refer to different masses as in the upper panel.}
\label{inertia_evol}
\end{figure}

According to this scenario, the initial  rotation period of the star $P_{\rm rot}$ is approximately equal to the orbital period of the planet, i.e., it is between 3 and 10 days. 
{ The typical lifetime of the disc does not exceeds
$5-10$ Myr which is shorter than the timescale of contraction to reach the zero-age main sequence (hereinafter ZAMS) for stars having a mass lower than $\approx 2 M_{\odot}$ \citep{Tinkeretal02,Bouvier08,Mamajek09}. When the disc disappers, stellar rotation is no more locked and the rotation period decreases during the approach to the ZAMS owing to the reduction of the moment of inertia of the star \citep{Scholzetal07,IrwinBouvier09}. In Fig.~\ref{inertia_evol}, we plot the evolution of the radius and the moment of inertia during the PMS phase, according to \citet{Siessetal00}, for stars of $1.0$, $1.2$, and $1.4$ M$_{\odot}$, respectively. The decrease of the moment of inertia is greater than expected on the basis of the contraction of the radius because the internal structure changes also with an increase of the mass of the radiative core as the star approaches the ZAMS.  The  reduction of the moment of inertia occurring between disc decoupling and arrival onto the ZAMS is by a factor of $\sim 5$ if the disc lifetime is 5  Myr. This implies a remarkable acceleration of stellar rotation  which destroys any synchronization with the planetary orbit attained during the previous disc-locking phase. This would give  $n/\Omega \sim 0.2-0.3$ for ZAMS sun-like stars with hot Jupiters, for which case there is no evidence in our sample of transiting planets. Therefore, we conjecture that some process is at work to restore synchronization when a planet-harbouring  star is approaching the ZAMS. A candidate mechanism is  a magnetocentrifugal stellar wind, as suggested by \citet{Lovelaceetal08}. Considering a star which was released by its disc with a rotation period of 8 days, it would reach the ZAMS with a period of only 1.6 days, if the reduction of the moment of inertia is not counteracted by any other process.  Assuming that the young contracting star has a surface magnetic field  of $10^{3}$~G, the torque exerted by its coronal field on the planet would transfer most of the stellar angular momentum to the planet itself on a time scale of $3-5$ Myr, restoring a synchronous rotation state. 
Recently, \citet{Vidottoetal09} have revisited such a mechanism considering a more realistic  wind model than the Weber \& Davis model adopted by \citet{Lovelaceetal08}. They find timescales longer by one order of magnitude for the angular momentum exchange between the star and the planet, which are still acceptable in the framework of our model. We conclude that a magnetocentrifugal wind may maintain synchronization in solar-like stars accompanied by a hot Jupiter during PMS evolution after the star has been released by its disc. This implies that the star arrives on the ZAMS in an approximate synchronous state of rotation. After the star has settled on the ZAMS, the efficiency of the stellar  hydromagnetic dynamo decreases with respect to its PMS phase because the volume of the outer convection zone is significantly smaller than in the PMS phase, so the magnetic field intensity at the surface drops and the coupling provided by the  magnetocentrifugal wind virtually vanishes. From this point on, the evolution of the spin and the orbital angular momentum  are decoupled and we can study the evolution of stellar rotation treating the angular momentum loss from the corona by means of the model of Sect.~\ref{angular_mom}. 

In addition to the scenario proposed above, another evolutionary sequence is possible 
if the magnetic field of the star  truncates the disc and couples the rotation of the star to its inner edge, but it is not strong enough to halt the migration of the planet (i.e., $\beta > 10-100$). In this case, the planet will continue to migrate inward until its orbital period becomes half of the period at the corotation radius because the angular momentum exchange between the planet and the disc proceeds via the 2:1 resonance  \citep[see  ][ and references therein]{Linetal06}. In this case, the initial rotation period of the star is  twice the orbital period of its hot Jupiter. If the star is massive enough, say at least $M \approx 1.5$ M$_{\odot}$, and its disc is long-lived, say, $\approx 10-15$ Myr \citep{Mamajeketal02}, it can reach the ZAMS while still being locked to its disc, thus starting its evolution in a rotational status with $n / \Omega \simeq 2$. If those stars do not appreciably loose angular momentum during their main-sequence evolution (see Sect.~\ref{ms_evol}), this may explain the observed concentration of systems with $T_{\rm eff} \ga 6200$ K around $n/ \Omega =2$ seen in Figs.~\ref{teff_corr} and~\ref{synch_vs_prot}.  
}

\subsection{Main-sequence evolution}
\label{ms_evol}

{ Starting from the initial status on the ZAMS described in Sect.~\ref{pms_evolution}, we want to account for the main features of the 
$n/\Omega$ distribution found in Sect.~\ref{observations}, namely the  dependence of $n/ \Omega$ on the effective temperature, with stars having $T_{\rm eff} \ga 6000$~K showing a generally smaller $n /\Omega$ than cooler stars, and the dependence of $n/ \Omega$ on the stellar rotation period,  found in stars having $T_{\rm eff} \ga  6000$~K. 

 To study  the angular momentum evolution on the main sequence, we  apply the model of Sect.~\ref{angular_mom}. The force-free parameter $\alpha$ of the coronal field plays a crucial role in that model.
}
\citet{Lanza08} proposed a method to estimate $\alpha$ in stars showing  chromospheric hot spots rotating synchronously with their hot Jupiters \citep{Shkolniketal05,Shkolniketal08}. To date, only five stars have been modelled, so conclusions based on such a method are still preliminary. Nevertheless, for F-type stars, i.e., \object{HD~179949}, \object{$\upsilon$~Andromedae}, and $\tau$ Boo, having $T_{\rm eff} > 6200$~K and $P_{\rm rot} < 12$ days \citep[cf. ][]{Shkolniketal08}, the values of $|\alpha|$ fall between 0.1 and 0.2, while for the two K-type stars \object{HD~189733} and \object{HD~192263}, having $T_{\rm eff} \simeq 5000$~K and $P_{\rm rot} > 12$ days \citep[cf. ][]{Santosetal03}, $|\alpha|$ ranges between 0.025 and 0.1.
Note that such values were obtained with the non-force-free model of \citet{Neukirch95}, but the typical values of $\alpha$ obtained with a purely force-free model do not differ by more that $10-20$ percent. 

A motivation for a greater value of $|\alpha |$ in  F stars than in  G and K stars may be the stronger  toroidal field  at their surface produced by a greater  relative differential rotation. Assuming that the measured photospheric field is a good proxy for the field at the base of the corona, an estimate of $\alpha$  can be obtained by comparing the first and the third of Eqs.~(\ref{BC_eqs}) which yields:
\begin{equation}
\alpha = - \frac{\int_{\Sigma(R)} B_{\phi}^{(s)} \sin \theta \, d \Sigma}{ R \int_{\Sigma(R)} B_{\rm r}^{(s)} \cos \theta \, d \Sigma}.  
\label{alpha_determ}
\end{equation}
Indeed spectropolarimetric observations of $\tau$ Boo by \citet{Donatietal08} and \citet{Faresetal09} show that the star has an oscillating field with a predominantly toroidal component during a significant fraction of its activity cycle. This may be a good example of a fast rotating F-type star with a hot Jupiter because its mean rotation period is $3.3$ days and it also shows a surface differential rotation with a relative amplitude of $\sim 0.2$ between the equator and the pole \citep{Catalaetal07}. 
In the case of CoRoT-4a, time series spot modelling suggests a  surface differential rotation comparable to that of $\tau$ Boo, thus supporting the presence of a predominantly toroidal surface field \citep{Lanzaetal09b}.

\citet{Barnesetal05} and \citet{Reiners06} show that the amplitude $\Delta \Omega$ of the surface differential rotation decreases strongly with the decrease of the effective temperature of the star, viz. $\Delta \Omega \propto
T_{\rm eff}^{8.9 \pm 0.3}$. Recent spectropolarimetric observations by \citet{Petitetal08} indicate that the photospheric magnetic field of G-type stars ($T_{\rm eff} \simeq 5700-6000$~K) with a rotation period below $10-12$ days is  predominantly toroidal, while stars with a longer rotation period have a predominantly poloidal field. Such a contrast may come from a different amplitude of the shear at the boundary between the radiative core and the convective envelope { (also called the tachocline in the Sun)}, which may be greater in hotter and fast-rotating stars. \citet{Bouvier08},  specifically considering stars with massive planets, suggested that their lower Lithium abundance may be the result of an enhancement of the turbulence at the core-envelope interface induced by hydrodynamic or magnetohydrodynamic instabilities associated to a sizeable shear localized at the interface. This suggests that a sizeable toroidal field is present in such stars, 
at least during the first phase of their evolution on the main sequence, 
produced by the shearing of a radial poloidal field close to the base of their convection zones. 

Considering  rapidly rotating ($P_{\rm rot} \leq 10-12$ days) F-type stars, we adopt $ 0.1 \leq |\alpha | \leq 0.2$, yielding a typical moment of inertia of their coronae ranging from $\sim 6 \times 10^{34}$ to $ \sim 10^{36}$ kg~m$^{2}$   (cf. Fig.~\ref{mom_i}).
The  timescale for angular momentum loss can be estimated as:
\begin{equation}
\tau_{\rm AML} = \frac{I_{*}}{I} \Delta t = \frac{\gamma^{2} M R^{2}}{I} \Delta t,
\label{aml_timescale}
\end{equation}
where $I_{*}= \gamma^{2} M R^{2}$ is the moment of inertia of the star, $\gamma R \sim 0.35 R$ is  its gyration radius, and $\Delta t$ is the mean time interval between the CME events that produce the loss of the  angular momentum of the stellar corona. 

We can estimate a lower limit for $\Delta t$ from the ratio between the total energy of the coronal field and the X-ray luminosity of the star, i.e., $\Delta t \approx E/L_{\rm X}$. For $\tau$ Boo and HD~179949, the average value of $L_{\rm X}$ is $\sim 3 \times 10^{21}$~W, while $E=7.6 \times 10^{27}$~J at the Aly limit for $B_{0}=10$ G \citep[a field intensity  measured in $\tau$~Boo by ][]{Donatietal08,Faresetal09}, giving $\Delta t \sim 2.6 \times 10^{6}$~s. An upper limit may come from the timescale for changing the global coronal field topology, as discussed in \citet{Lanza09}, i.e., $\Delta t \sim 2.6 \times 10^{7}$~s, or $\sim 300$ days. We shall adopt $\Delta t = 2.6 \times 10^{6}$~s because a variation of $\Delta t$ in Eq.~(\ref{aml_timescale}) can be compensated by a change of $I$ given that the density $\rho_{0}$ at the base of the corona may vary by one order of magnitude.  

In the case of $\tau$~Boo, $R=1.6$~R$_{\odot}$, $M =1.2$~M$_{\odot}$, and 
$| \alpha | = 0.12$, yielding $I \sim 3 \times 10^{35}$~kg~m$^{2}$; thus we find an angular momentum loss timescale of the order of $\tau_{\rm AML} \approx 100$~Gyr. Such a value implies that the initial angular momentum of $\tau$~Boo remains approximately constant during its main-sequence lifetime. In other words, the observed  synchronization  between the average stellar rotation and the orbital period of the planet should be a remnant of the initial state of the system when the star settled on the ZAMS. A similar conclusion is reached for CoRoT-4 
\citep{Lanzaetal09b}. An angular momentum loss time scale of the order of 100 Gyr
accounts also for the rotation periods of the mid-F type stars in the systems XO-4 and HAT-P-6, which again appear to be remnants of their ZAMS rotational status, in this case with an initial  $n / \Omega \simeq 2$. 

{ In the light of the results of \citet{Petitetal08}, stars with $T_{\rm eff} \geq 6000$~K and rotation periods longer than $\sim 10$ days should be characterized by a smaller  surface toroidal field than more rapidly rotating stars which implies a smaller value of $\alpha$. Therefore, their angular momentum loss time scale is expected to be shorter than that of the rapidly rotating F-type stars considered above, which may account for the dispersion of $n /\Omega$ observed in the effective temperature range $6000-6500$~K. For systems such as WASP-1, WASP-15, or WASP-18, adopting 
$R=1.4$ R$_{\odot}$, $M=1.2$ M$_{\odot}$, and $|\alpha|=0.08$, we have $I \sim 6 \times 10^{36}$ kg m$^{2}$, so we  find $\tau_{\rm AML} \sim 4$ Gyr for $\Delta t \sim 2.6 \times 10^{6}$~s. Therefore, the scatter in $n / \Omega$ observed for $P_{\rm rot} > 10$ days may be explained as a consequence of the different stellar ages.  Note also that for WASP-18 the high value of $n / \Omega$ can be due to the very short orbital period which results from a very strong tidal interaction in a regime with $ n > \Omega$ \citep[cf. ][]{Hellieretal09a}.  
}  

In the case of stars of spectral types G and K we assume that the value of $|\alpha|$ is significantly lower than in the case of F-type stars. This is justified because their differential rotation is lower than that of hotter stars, given the remarkable dependence of $\Delta \Omega$ on $T_{\rm eff}$. In turn, this implies a lower toroidal field yielding a lower  $\alpha$ at the same rotation period.
Considering a mean value of 
$|\alpha| \sim 0.06$, we have a  coronal moment of inertia $I \sim 2 \times 10^{37}$ kg m$^{2}$. For a star with the mass and the radius of the Sun, with 
$\Delta t \sim 3 \times 10^{6}$ s, this implies $\tau_ {\rm AML} \approx 500$ Myr. Such a timescale corresponds to that of the initial fast angular momentum loss occuring on the main sequence during the transition between the two braking sequences introduced by \citet{Barnes03}, i.e., from the so-called convective to the interface  sequence. Note that for a mid-G-type star without a close-in planet such a transition occurs in $100-300$ Myr. Therefore, the effect of a hot Jupiter is that of slowing down the initial angular momentum evolution of G stars by a factor of $\sim 2-5$. The same is true also for K-type stars, but, since their transition from the convective to the interface sequence takes longer ($ \approx $ 500-800 Myr), the effect of the close-in planet is less important. 
 
{ In conclusion, in the case of a G- or K-type  star accompanied by a hot Jupiter, we expect a significant slowing down of the initial phase of its  rotational braking, particularly when its initial rotation period is shorter than $8-10$ days and the star has a sizeable photospheric azimuthal field component. 
When its rotation period becomes longer than $\approx 10$ days, its toroidal field component declines steeply 
\citep[cf., ][]{Petitetal08}
leading to a decrease of $|\alpha|$ and a remarkable increase of the angular momentum loss rate. In this phase, the rate of angular momentum loss might become similar to that of stars without planets
 and the subsequent evolution could not be remarkably affected by the presence of a hot Jupiter, i.e., the star would  continue to spin down according to the usual Skumanich law characteristic of stars on the so-called interface sequence of \citet[][ cf. Sect.~\ref{observations}]{Barnes03,Barnes07}. Considering the different ages of planet-harbouring stars and their different initial rotation periods, we may explain the larger dispersion of $n / \Omega$ observed in stars with $T_{\rm eff} \la 5800$~K (cf. Fig.~\ref{teff_corr}).

In our treatment of the main-sequence spindown we have assumed that a star is braked as a rigid body (cf.~Eq.~\ref{aml_timescale}). This hypothesis is adequate in the present case because our braking time scales $\tau_{\rm AML}$ are generally  longer than the time scale for angular momentum exchange between the radiative interior and the outer convection zone which evolutionary models of stellar rotation set at $\la 100$ Myr  on the main sequence \citep[cf. ][]{Bouvier08,IrwinBouvier09}. For the same reason, the tidal synchronization time should be computed by considering the spin-up of the whole star, as we did in Sect.~\ref{observations}, not just of its convection zone. 
}

\subsection{A tentative comparison with observations }
{ In the framework of a Skumanich-type braking law, \citet{Barnes07} provides an empirical formula to estimate the age of a main-sequence star from its rotation period and colour index. We apply it to HD~149026, HAT-P-1 and WASP-15 to test the predictions of our model for stars with a rotation period $P_{\rm rot} > 10$ days and $6000 < T_{\rm eff} < 6300$~K. These three systems have been selected because they have a tidal synchronization time at least 3 times longer than their maximum estimated ages, in order to exclude tidal effects on their angular momentum evolution. Their ages, as estimated with Eq.~(3) of \citet{Barnes07} are 2.2, 2.0 and 6.0 Gyr, respectively. They are all within the range of ages estimated by isochrone fitting, as reported in Table~\ref{table_p2}. For the first two stars, the gyrochronology ages are close to the lower limit given by isochrone fitting, while for WASP-15, the gyro age is close to the isochrone upper bound. Therefore, this  preliminary comparison suggests that some reduction of the angular momentum loss rate may still be induced by a  close-in massive planet when $P_{\rm rot} > 10$ days and $6000 < T_{\rm eff} < 6300$ K, at least in some cases, although this needs to be confirmed by a larger sample of systems. { Note that in Sect.~\ref{observations} we found a similar result based on a greater sample of stars providing us with  significant statistics. However, in that case we took into account the evolution of  angular momentum and the tidal effects in separate analyses  in order to have a significant sample in both cases. Now, we have considered the evolution of the angular momentum of stars selected to have negligible tidal effects, which severely restricts our sample.  } 

A major limitation of the present approach is that stellar ages derived from  isochrone fitting are highly uncertain, especially for stars with $M \leq 1$ M$_{\odot}$. Therefore, better age estimates are needed, such as those derived for open cluster members. Searches for transiting planets in open clusters have just begun and it is hoped that they may contribute to clarify this issue \citep[e.g., ][]{Montaltoetal07,Hartmanetal09a}.   
}

\section{Conclusions}

{ We have analysed  the rotation of stars harbouring transiting hot Jupiters and have found a general trend  toward synchronization with increasing effective temperature. Stars with 
$T_{\rm eff} \geq 6500$~K are synchronized or have a rotation period close to twice the orbital period of their planets ($n / \Omega \simeq 1$ or $2$, respectively), while those with $6000 < T_{\rm eff} < 6500$~K have $n/\Omega \simeq 1$ or $2$ only for $P_{\rm rot} < 10$ days. Stars with $T_{\rm eff} \la 6000$~K generally show rotation period remarkably longer than the orbital periods of their planets. 

We conjecture that planet-harbouring stars are borne with circumstellar discs  in which hot Jupiters form and migrate inward while the disc  locks the rotation of the star. Depending on the magnetic field strength in the inner region of the disc, two different migration scenarios are possible, leading to a state with $n/ \Omega \simeq 1$ or $n /\Omega \simeq 2$, respectively. When the discs disappear, most of the stars with $M \leq 1.5$ M$_{\odot}$ are still contracting toward the ZAMS, so their rotation accelerates owing to the reduction of their moment of inertia. Nevertheless, we conjecture that the synchronization between stellar rotation and planetary orbit is maintained throughout  the final phases of the PMS evolution by the strong coupling provided by a magnetocentrifugal stellar wind \citep{Lovelaceetal08}. Stars with $M \geq 1.4-1.5$ M$_{\odot}$ and very long lived discs ($\approx 15$ Myr) may arrive on the ZAMS while still locked to their discs, thus starting their main-sequence evolution in a status with $n /\Omega \simeq 2$. }

Once a star has settled on the ZAMS, its rotational evolution is ruled by the angular momentum loss from its corona. 
We assume that stars 
accompanied by close-in giant planets  have a coronal magnetic field dominated by closed field lines, so that most of their angular momentum loss occurs  through eruptions similar to the solar coronal mass ejections rather than via a continuously streaming stellar wind. 
{ This peculiar  configuration is induced by the steady motion of the planet through the stellar corona which reduces the magnetic helicity of the coronal field leading to a predominance of closed magnetic loops \citep[e.g., ][]{Lanza09,Cohenetal09}. 
}
Using a simple linear force-free field, we estimate the angular momentum loss rate for different field geometries characterized by different values of the force-free parameter $\alpha$. We find that the angular momentum loss decreases by two orders of magnitude when $| \alpha |$ ranges from $\sim 0.05$ to $\sim 0.2$. 

If  $| \alpha | \sim 0.15-0.2$ is characteristic of F-type stars with $T_{\rm eff} \ga 6000$~K and ZAMS rotation periods $P_{\rm rot} \la 10$ days, their rotational evolution  requires timescales of the order of $30-100$ Gyr, that is those planetary systems would be characterized by an almost constant distribution of spin and orbital angular momentum all along their main-sequence lifetime, with their present status reflecting their  angular momentum distribution on the ZAMS. 

On the other hand, F-type stars with a rotation period initially longer than $\sim 10$ days are characterized by
a  smaller value of $\alpha$, say, $\sim 0.08-0.1$, leading to a greater angular momentum loss during coronal mass ejections. Their  spin is expected to evolve on a timescale of $\sim 4-7$ Gyr, leading to some spreading in the distribution of $n/ \Omega$ in the effective temperature range $6000-6500$~K as a consequence of the different ages of the stars.
Later-type stars are characterized by still smaller values of $\alpha$, i.e., $\sim 0.05$, leading to shorter braking time scales. 
Therefore, the angular momentum evolution of planet-harbouring G- and K-type stars should not be dramatically different from that of stars without close-in massive planets { (cf. Sect.~\ref{observations})}. However, a reduction of the angular momentum loss rate by a factor of $2-5$ may still be caused by planets around young, rapidly rotating ($P_{\rm rot} \la 10$ days) stars.

Such predictions can be tested 
by increasing the sample of F, G and K stars with known hot Jupiters, especially in open clusters of different ages allowing us to compare the rotational evolution of  coeval stars with and without close-in planets. However, since open cluster members are usually faint, this requires  dedicated programs to be conducted with large telescopes to reach the necessary photometric and radial velocity precisions. 

Asteroseismology can provide stellar ages with an accuracy of $\approx 10$ percent of the total stellar  main-sequence lifetime \citep[e.g., ][]{Kjeldsenetal09}, but the internal chemical composition of planet-hosting stars may  differ from that of their surface layers inducing systematic errors \citep[cf., e.g., ][]{BazotVauclair04}.

{ 
In principle, spectropolarimetric techniques can be applied to derive the value of the parameter $\alpha$ in stars harbouring hot Jupiters, provided that they rotate sufficiently fast (cf. Sect.~\ref{field_model}). This should allow us to test our theory in the case of individual objects, at least those with a sufficiently rapid rotation. 
}

The possible effect of hot Jupiters on stellar angular momentum loss must be taken into account when interpreting  the results  of \citet{Pont09} in the sense that they could not necessarily provide evidence that tides are ruling the spin evolution in stars with close-in planets. It is more likely that both tides and the effects discussed in this paper are simultaneously at work to affect the distribution of  angular momentum and its evolution in stars harbouring hot Jupiters. 

Finally, we note that gyrochronology may not be suitable to estimate the age of  late-type stars with close-in giant planets, especially if they have $T_{\rm eff} \geq 6000$~K and/or are  rotating with a period shorter than $\sim 10$ days, because their rotational evolution can be remarkably different from that of stars without hot Jupiters. 

\begin{acknowledgements}
The author is grateful to an anonymous Referee for a careful reading of the manuscript and several interesting and stimulating comments. 
Active star research and exoplanetary studies at INAF-Catania Astrophysical Observatory and the Department of Physics and Astronomy of Catania University is funded by MIUR ({\it Ministero dell'Istruzione, Universit\`a e Ricerca}), and by {\it Regione Siciliana}, whose financial support is gratefully
acknowledged. 
This research has made use of the ADS-CDS databases, operated at the CDS, Strasbourg, France.
\end{acknowledgements}

\begin{footnotesize}

\end{footnotesize}


\end{document}